\documentclass[a4paper,superscriptaddress,floatfix,nofootinbib,onecolumn,longbibliography,notitlepage,onecolumn]{revtex4-1}
\usepackage{bbm}
\usepackage{bbold}
\usepackage{bbm}
\usepackage[pdftex]{graphicx}
\usepackage{latexsym,amsmath,verbatim,amssymb,txfonts}
\usepackage{lipsum}
\usepackage{color}
\usepackage{rotating}
\usepackage{verbatim}
\usepackage{multirow}
\usepackage[english]{babel}
\usepackage{comment}
\usepackage{bm}
\usepackage{amsmath}
\usepackage{amsfonts}
\usepackage{amssymb}
\usepackage{float}
\usepackage{color}
\usepackage{braket}
\usepackage{blkarray, bigstrut}

\newcommand\id{\mathbb{I}}

\DeclareMathOperator{\lap}{\mathbf{L}}
\DeclareMathOperator{\bnd}{\mathbf{B}}

\begin{document}

\title{Turing patterns on discrete topologies: from networks to higher-order structures}

\author{Riccardo Muolo}
\email{muolo.r.aa@m.titech.ac.jp}
\affiliation{Department of Systems and Control Engineering, Tokyo Institute of Technology, Japan}
\affiliation{Department of Mathematics \& naXys, Namur Institute for Complex Systems, University of Namur, Belgium}

\author{Lorenzo Giambagli} 
\affiliation{Department of Mathematics \& naXys, Namur Institute for Complex Systems, University of Namur, Belgium}
\affiliation{Department of Physics and Astronomy, INFN \& CSDC, University of Florence, Italy}

\author{Hiroya Nakao}
\affiliation{Department of Systems and Control Engineering, Tokyo Institute of Technology, Japan}
\affiliation{International Research Frontiers Initiative, Tokyo Institute of Technology, Japan}

\author{Duccio Fanelli} 
\affiliation{Department of Physics and Astronomy, INFN \& CSDC, University of Florence, Italy}

\author{Timoteo Carletti}
\affiliation{Department of Mathematics \& naXys, Namur Institute for Complex Systems, University of Namur, Belgium}

\begin{abstract}
Nature is a blossoming of regular structures, signature of self-organization of the underlying microscopic interacting agents. Turing theory of pattern formation is one of the most studied mechanisms to address such phenomena and has been applied to a widespread gallery of disciplines. Turing himself used a spatial discretization of the hosting support to eventually deal with a set of ODEs. Such an idea contained the seeds of the theory on discrete support, which has been fully acknowledged with the birth of network science in the early 2000s. This approach allows us to tackle several settings not displaying a trivial continuous embedding, such as multiplex, temporal networks, and, recently, higher-order structures. This line of research has been mostly confined within the network science community, despite its inherent potential to transcend the conventional boundaries of the PDE-based approach to Turing patterns. Moreover,  network topology allows for novel dynamics to be generated via a universal formalism that can be readily extended to account for higher-order structures. The interplay between continuous and discrete settings can pave the way for further developments in the field.

 \end{abstract}

\maketitle

 \section{Introduction}

Patterns are all around us. From the synchronized firing of neurons to the spots and stripes on animal coats, the emergence of coherent spatio-temporal structures and self-organization phenomena seems ubiquitous in natural and engineered systems. One of the most popular theories explaining such ordered motifs is due to British mathematician Alan Turing.  {In a} milestone work published in 1952, {Turing} showed that, by perturbing a reaction-diffusion system {made of} two species in a stable homogeneous equilibrium (a fixed point), {one can trigger} a diffusion-driven {instability that takes} the system towards an asymptotically stable inhomogeneous state \cite{Turing}, i.e., a pattern. Successively, Gierer and Meinhardt realized that one of the two species needed to be an activator, while the other one an inhibitor \cite{GiererMeinhardt}. Nowadays, the theory is known as Turing instability and the consequent patterns are known as Turing patterns. Such a mechanism of pattern formation has found many applications in different domains: from neuroscience \cite{bressloff2002geometric} to genetics \cite{kondo2010reaction,economou2012periodic} and {   morphogenesis \cite{mizuguchi1995proportion}}, from economics \cite{helbing2009pattern} and linguistics \cite{mimar2021linguistic} to nano-scale systems \cite{turing_nano}, and even to quantum mechanics \cite{kato2022turing} {   and robotics \cite{slavkov2018morphogenesis}}. One issue with this remarkable theory is that the conditions to obtain such an instability are quite restrictive and, in fact, Turing patterns have been obtained experimentally only {  about 40 years} {after the original conception of the theory} and with great struggle \cite{Castets}; this is in contrast with the abundance of patterns observed in nature. Scholars have hence tried to relax some of the conditions for Turing instability{  . For example, in Tompkins et al., the authors introduced heterogeneity \cite{tompkins2014testing}, while} Ninomiya recently {showed} that patterns can {indeed} emerge via a Turing-like mechanism even when the diffusion coefficients of the two competing species are equal \cite{ninomiya2024example}.
Other endogenous and exogenous factors have been considered to generalize the theory beyond its original boundaries. {In Butler and Goldenfeld \cite{butler2009robust,butler2011fluctuation}, and  Biancalani et al. \cite{biancalani}, it was for instance shown that the stochastic setting is more prone to yield patterns. This remarkable observation has been invoked to explain the spontaneous drive to self-organization for a selected number of case studies, where demographic noise - hence, endogenous stochasticity - is at play \cite{goldenfeld, stavans, lavacchi, dipatti}. Furthermore,} Van Gorder studied the effects of temperature, showing that it acts as a sensible parameter to either repress or boost the onset of the instability \cite{van2020influence}. Other variations include non-local interactions \cite{ninomiya2017reaction}, cross-diffusion coefficients \cite{fanelli_cianci}, the presence of an upper bound in the signal propagation yielding to hyperbolic reaction-diffusion systems \cite{ZH2016} and the two combined effects \cite{CurroValenti}. In particular, the hyperbolic setting allows the emergence of oscillatory patterns \cite{ritchie}, which would not be permitted in the reference two-species settings. As Turing already remarked in his original paper, at least three {simultaneously interacting} species are in fact needed for the oscillation to manifest via the conventional recipe \cite{hata2014sufficient,anma_wave}.

All the above results {refer to models described in terms of Partial Differential Equations (PDEs), while still accounting for a large plethora of applications, ranging from} ecology to chemistry and neuroscience, to name a few, that can otherwise be considered as
intrinsically discrete {in space. As such,} they find a natural embedding onto a discrete support rather than a continuous one, {as customarily done when invoking the Turing instability route to pattern formation}. For instance, if one thinks of neuroscience, it is intuitive to visualize the system as a network, whether one considers single neurons connected by axons and dendrites, or different brain modules connected between each other. It is thus natural to wonder if the {foundational aspects of the Turing} theory can be extended to {account for this generalized} framework. Throughout this {Review paper}, we will show not only that the answer is positive, but also that the network approach allows us to treat in a simple way certain settings that would be extremely challenging or even impossible to deal with on continuous support. {  Moreover, from a network standpoint, it is natural to discuss as further extension the inclusion of higher-order interactions, opening up new and exciting possibilities for revisiting Turing theory from a modern perspective}.

We intend this Review as complementary to a recent special issue "Recent progress and open frontiers in Turing’s theory of morphogenesis" \cite{KGMK_intro2021}, which provides interesting insights {to the contributions in the frame of {  PDEs} modeling, that has been routinely advocated in} Mathematical Biology. On the other hand, a Review on the recent progress of the {Turing theory adapted to} networked systems is lacking. Hence, our aim is two-fold. On the one side we shall review the main results {as reported in the literature}, moving from the applications of Turing theory on networks to cover the more {recent generalization to the growing field of higher-order interactions}. On the other side, we will introduce readers from the PDEs community to the {widespread realm of network science} and guide them through the advantages that it offers. Our hope is to foster more interactions between the PDEs and network communities, as they would both benefit from their complementary expertise in further developing Turing theory {as a seminal mechanism to spontaneous} pattern formation. Given such a double purpose, this work will alternate review parts and pedagogical sections. In Sec. \ref{sec:1st_approach}, we will introduce the extension of Turing theory to discrete domains following the discretization approach, which led to the modern version of the theory on complex networks. A review of the main results of Turing theory on networks can be found in Sec. \ref{sec:extensions_network}, where some specific settings not having trivial continuous analogues will be discussed in detail. In particular, we will focus on directed and non-normal networks, multiplex and, lastly, temporal networks. {   Then, in Sec. \ref{sec:2nd_approach}, we will show how the above formalism can be extended to oscillatory systems, in the framework of synchronization theory.} Moreover, and as stated earlier, we will make contact with the area of higher-order interactions, a newly established field of investigation that holds promise for exciting developments. {   Concretely, Sec. \ref{sec:higher} will be dedicated to the extension of the Turing framework for systems with higher-order interactions} and, lastly, in Sec. \ref{sec:top}, we will go even further by considering reaction-diffusion systems of state variables of different dimensions, called \textit{topological signals}. In the last Section, we will sum up and provide an up to date list of open problems and possible directions for future investigations. 

In what follows, we will assume the reader to be already familiar with the main aspects of Turing theory on continuous support. For those who are not, we suggest going through the basics of the theory before proceeding any further: besides Turing's original paper \cite{Turing}, a good pedagogical introduction can be found in \cite{Murray2001}. On the other hand, we will introduce the main concepts of network theory, higher-order interactions and topological signal theory, making this Review self-consistent for readers from the PDEs community. In particular, {   more details will be given in the part on higher-order interactions, Secs. \ref{sec:higher} and \ref{sec:top}, due to the novelty of the topic}.

\section{Turing theory on networks: discretization of continuous domains approach}\label{sec:1st_approach}

A natural way to think of a network is to consider a discretization of a continuous manifold. For simplicity, let us consider a $1$-dimensional domain with periodic boundary conditions, i.e., a circle, whose discretization returns a $1$-dimensional lattice, i.e., a ring {  network}. Observe that this approach has been widely used when considering meshes to study systems otherwise defined on continuous support. Indeed, Turing himself in his original paper considered the reaction-diffusion problem on a ring of cells~\cite{Turing}. $20$ years later,  two complementary branches of the theory emerged, one that deals with systems hosted on a continuous support and the other that focuses on networks: in fact, Gierer and Meinhardt developed Turing's idea in the framework of PDEs \cite{GiererMeinhardt}, while Othmer and Scriven studied pattern formation in cellular networks \cite{OS1971,OthmerScriven74}. The latter works, involving various regular lattices as physical support for the reaction and diffusion process, are considered the firsts to address the problem on a discrete support, since Turing's pioneering paper. After about $20$ years, Goraş, Chua and collaborators were able to design Cellular Neural Networks capable of reproducing Turing patterns \cite{goras1995a,goras1995b,goras1995c}. The above works, despite their seminal approach, did not fully explore the possibilities offered by the network framework, because this research domain was still in its infancy. In fact, only in the early 2000s scholars realized the power and ductility of the network approach, which was made popular by some important works by Albert and Barab\'asi \cite{albert2002statistical}, Newman \cite{newman2002spread}, Strogatz \cite{Strogatz01Nature}, Pastor-Satorras and Vespignani \cite{pastor_vesp2001}, to name a few. Horsthemke and Moore made use of a discrete topology to model arrays of coupled chemical reactors in \cite{horsthemke2004}, and then, together with Lam, in \cite{horsthemke2004b}. The latter is the first work in which a non-regular discrete structure is being considered. Let us observe that, despite the novelty, those works did not fully exploit the complex network features: indeed, the size of the considered networks was small and no relation has been established between the topology of the hosting network and the employed dynamical system. However, all the above works set the basis on which Turing theory was finally extended in its modern version on complex networks by Nakao and Mikhailov in \cite{NM2010}. In what follows, we are going to provide a pedagogical introduction to Turing theory on networks, starting with an introduction of the network framework in Sec.~\ref{sec:1st_approach}\ref{sec:networks}, to then go through the analysis carried out by Othmer and Scriven \cite{OS1971} and Nakao and Mikhailov~\cite{NM2010}.

\subsection{The network framework and the discrete Laplacian}\label{sec:networks}

To introduce Network Theory in one paragraph is not an easy task, and, in fact, we do not aim to fulfill it here. There are several excellent books devoted to this topic, e.g., \cite{newmanbook2,latora_nicosia_russo_2017,barabasibook}, where the interested reader can find additional information and deepen their understanding. In this work, we will mainly focus on dynamics on networks, rather than network theory \textit{tout court}; hence, we will {limit ourselves} to introduce only the concepts which are fundamental for the specific framework to be hereby addressed.

A network (or graph) is an ordered couple $\mathcal{G}=(V,E)$ where $V$ is the {  node (vertex) set} and $E$ is the {  link (edge) set}.  Let us denote by $(i,j)$ a direct link from node $i$ to node $j$, and define thus node $i$ to be adjacent to node $j$. In a broad perspective, this encapsulates the fact that node $i$ exerts some impact onto node $j$. The resulting structure will be a directed network. If, on the other hand, for all pair of nodes $i$ and $j$ the existence of the link  $(i,j)\in E$ implies also the existence of the reciprocal one, $(j,i)\in E$, then the network is undirected (or symmetric). Except for Sec. \ref{sec:extensions_network}\ref{sec:dir_nn}, unless otherwise specified, we will only deal with undirected networks. 

A directed network of $n$ nodes can be represented in terms of its {adjacency matrix}, $A$,
\begin{displaymath}
\label{adj2}
A_{ij}=\begin{cases} 1 ~~~\mbox{if  }(j,i)\in E, \\ 0~~~\mbox{otherwise}, \end{cases} ~~~~ \forall i,j=1,....,n \, .
\end{displaymath}
The above is thus an asymmetric {square matrix}, whose number of rows (and columns) is the number of nodes that the network encodes. If the examined network is undirected,  the associated matrix $A$ is symmetric, because the existence of the link $(j,i)$, $A_{ij}=1$, implies the existence of the reciprocal one, thus $A_{ji}=1$. We will  {denote by} $\{i,j\}$ a symmetric link between nodes $i$ and $j$. Networks can in general be weighted, i.e., $A_{ij}\geq 0$. In principle, there could be {  other features, such as} negative weights or self-loops, but we will not consider any of the above cases here, for the sake of simplicity. The interested reader may consult one of the aforementioned textbooks to elaborate further on these points.

{An important feature associated to each node $i$} is {its} degree, namely, the number of nodes {adjacent} to it. If the network is undirected, we define $k_{i}$ the degree of node $i$ as: 
\begin{displaymath}
 k_{i}=\sum_{j=1}^{n}A_{ij},~~~~~~~\forall i=1,....,n,
\end{displaymath}  while, if it is {  directed}, we need to distinguish between out-going and in-coming connectivity: 
\begin{displaymath} 
\displaystyle k_{i}^{out}=\sum_{j=1}^{n}A_{ji} \quad \text{  and  } \quad \displaystyle k_{i}^{in}=\sum_{j=1}^{n}A_{ij}, \quad \forall i=1,....,n.
\end{displaymath}  

Let us now discuss the process of diffusion on networks following a heuristic argument, as in \cite{newmanbook}. {   We start from an undirected network of $n$ nodes and consider a given substance (or species) $u$, whose concentration at node $i$ is denoted by $u_i$. Now, let us focus on  the concentration on node $j$, $u_j$. The species on node $j$ relocates towards node $i$ following the symmetric link $\{i,j\}$, with rate $D_u(u_{j}-u_{i})$, where $D_u$ stands for the diffusion coefficient of the examined substance. In a small interval of time $dt$, the amount of substance that goes from node $j$ to node $i$ is $D_u(u_{j}-u_{i})dt$. By taking into account all nodes $j$ connected to node $i$, one can write:
\begin{displaymath}
\dot{u}_{i}=D_u\sum_{j=1}^{n} A_{ij}(u_{j}-u_{i})=D_u\sum_{j=1}^{n}A_{ij}u_{j}-D_u u_{i}\sum_{j=1}^{n} A_{ij}=D_u\sum_{j=1}^{n} A_{ij}u_{j}-D_u u_{i}k_{i},
\end{displaymath} 
i.e., the substance entering node $i$ from the adjacent nodes $j$, minus the substance leaving node $i$, as stipulated by the Fick's law. By defining $L_{ij}=A_{ij}-\delta_{ij}k_{i}$, we obtain the equation describing the diffusion process, namely, \begin{displaymath}
\dot{u}_{i}=D_u\sum_{j=1}^{n}L_{ij}u_{j}.
\end{displaymath}
The matrix $\mathbf{L}$ of entries $L_{ij}=A_{ij}-\delta_{ij}k_{i}$ is termed (combinatorial) {\em Laplacian} \cite{lambiotte_book}. In Section~\ref{sec:top}, we will provide an alternative construction of the Laplacian matrix allowing to draw a formal connection between networks and simplicial complexes. Such a matrix reflects the topology of the support, i.e., it is symmetric (resp. asymmetric) when the network is undirected (resp. directed).} We can observe that the Laplacian matrix replaces the second-order differential operator used in the case of continuous domains; indeed, in the case of a regular lattice, {   the matrix $\mathbf{L}$ returns the usual finite difference approximation of the relevant continuous differential operator. While discretizing exotic continuous manifolds proves in general difficult \cite{macdonald2010implicit,woolley2017turing}, the combinatorial Laplacian operator $L$ can be straightforwardly implemented as long as the network topology, i.e., the adjacency matrix $\mathbf{A}$, is known. This observation testifies on the inherent flexibility of the network approach to tackle the modeling of dynamical systems.} For undirected networks, the Laplacian matrix is symmetric and, by construction, it has non-positive real eigenvalues, $\Lambda^{(\alpha)}$. Moreover, one can prove that the largest eigenvalue always vanishes,  $\Lambda^{(1)}=0$, and, for a connected network, its associated eigenvector is given by $(1,\dots,1)^\top$.

Before concluding, let us point out that the above expression for the Laplacian matrix is sometimes defined with a minus sign, i.e., $\delta_{ij}k_{i}-A_{ij}$, meaning that the matrix is now positive semi-definite. Such definition is mainly found in works dealing with synchronization \cite{arenas2008synchronization}, but also in mathematical graph theory and, hence, in some of the reference textbooks, e.g., \cite{newmanbook2}. In the Turing community, it is customary to stick to the negative semi-definite formulation of the Laplacian matrix, because of the analogy with the continuous diffusion operator. Moreover, the combinatorial Laplacian is not the only possible network Laplacian. Other options are the random walk Laplacian \cite{masuda2017} and the consensus Laplacian \cite{CENCETTI2020109707}. {  A fundamental difference regards the spectrum}, which is normalized and, thus, bounded between $0$ and $-2$ in the latter cases, while it takes values between $0$ and $-2k_{max}$ (negative twice the maximum degree) for the former. In the following, we will 
 refer to the combinatorial negative semi-definite Laplacian $L_{ij}=A_{ij}-\delta_{ij}k_{i}${  , except for Sec. \ref{sec:top}, where the Laplacian will be positive semi-definite.}

\subsection{Derivation of the Turing instability theory}\label{theory_derivation}

As discussed in the Introduction, Turing's brilliant idea consists in considering a system of two interacting species {exhibiting} a homogeneous equilibrium {that can become unstable under heterogeneous perturbations. Because the diffusive part of the system is responsible for this symmetry breaking, the Turing instability is also named {\em diffusion-driven instability}}. {  This} latter seeds an exponential growth of the {heterogeneous} perturbation, which takes the system away from the homogeneous equilibrium {to possibly achieve} a new inhomogeneous one, i.e., a pattern. {In the following, we will consider a metapopulation approach \cite{May}: species are anchored to nodes that are considered as well-stirred reservoirs. The local reactive dynamics is hence governed by a set of ordinary differential equations (i.e., mean-field). Then, as previously described, the diffusion process allows species to move across nodes following Fick's law and no reactions occur on the links}. Such a setting can be visualized in Fig. \ref{fig:1}(a). {To provide a self-consistent interpretation of the scrutinized phenomenon, but also to avoid cumbersome computations, we will hereafter solely focus on the main steps of the derivation, highlighting the differences between  continuous and {discrete} frameworks. The interested reader can find a step-by-step derivation of the conditions for {the emergence of} the Turing instability in the Methods section of {   Ref.} \cite{NM2010}. Let us start from the equations describing an isolated system composed of two species $u$ and $v$, given by the following nonlinear equations:
\begin{equation}\label{eq:react}
    \begin{cases} \dot{u}(t)=f(u(t),v(t)), \\
    \dot{v}(t)=g(u(t),v(t)),
    \end{cases} 
\end{equation}
and assume that there exists a stable equilibrium point $(u^*,v^*)$, i.e., $f(u^*,v^*)=g(u^*,v^*)=0$. Following the {analysis provided by} Gierer and Meinhardt \cite{GiererMeinhardt}, we consider $u$ as the activator and $v$ as the inhibitor. This means that $u$ is self-producing and activates the production of $v$, i.e., the derivatives of $f$ and $g$ with respect to $u$ at the equilibrium are positive, while the $v$ is self-annihilating and inhibits the production of $u$, i.e., the derivatives of $f$ and $g$ with respect to $v$ at the equilibrium are negative. We then consider the above system {defined} on a complex network~\cite{OS1971,NM2010}. We thus assume to deal with $n$ identical copies of~\eqref{eq:react}, each one anchored to a node where the reaction occurs, i.e., we assume to have well-stirred concentrations, and, by invoking Fickean diffusion, we allow chemicals to move through existing links. When the system is {at the} homogeneous equilibrium, i.e., $u_i=u^*$ and $v_i=v^*$, there is no concentration gradient {and the Fick contribution thus vanishes. Observe that this can be reformulated by saying that the} nodes do not ``feel'' the presence of each other and their dynamics can be considered as isolated. This is why in the network framework we call such a system \textit{isolated}, rather than \textit{aspatial} as in the PDEs framework. With all the above, the form of reaction-diffusion equations in this setting is \begin{equation}\label{eq:react_diff_eq}
    \begin{cases} \dot{u}_i=f(u_i,v_i)+D_u \displaystyle\sum_{j=1}^n L_{ij} u_j, \\
    \dot{v}_i=g(u_i,v_i)+D_v \displaystyle\sum_{j=1}^n L_{ij} v_j,
    \end{cases} \forall i=1,2,\dots,n.
\end{equation}

{Because $\sum_j L_{ij}=0$ and we assume nodes to bear the same reaction terms} (i.e., $f$ and $g$ are independent of the node), $(u^*,v^*)$ is also a solution of the coupled system. {To study the stability of this global solution $(u_i,v_i)=(u^*,v^*)$ for all $i,j$, we can perform a linear stability analysis about the latter, by obtaining for every node $i$, 
\begin{figure}
\centering
\includegraphics[width=\textwidth]{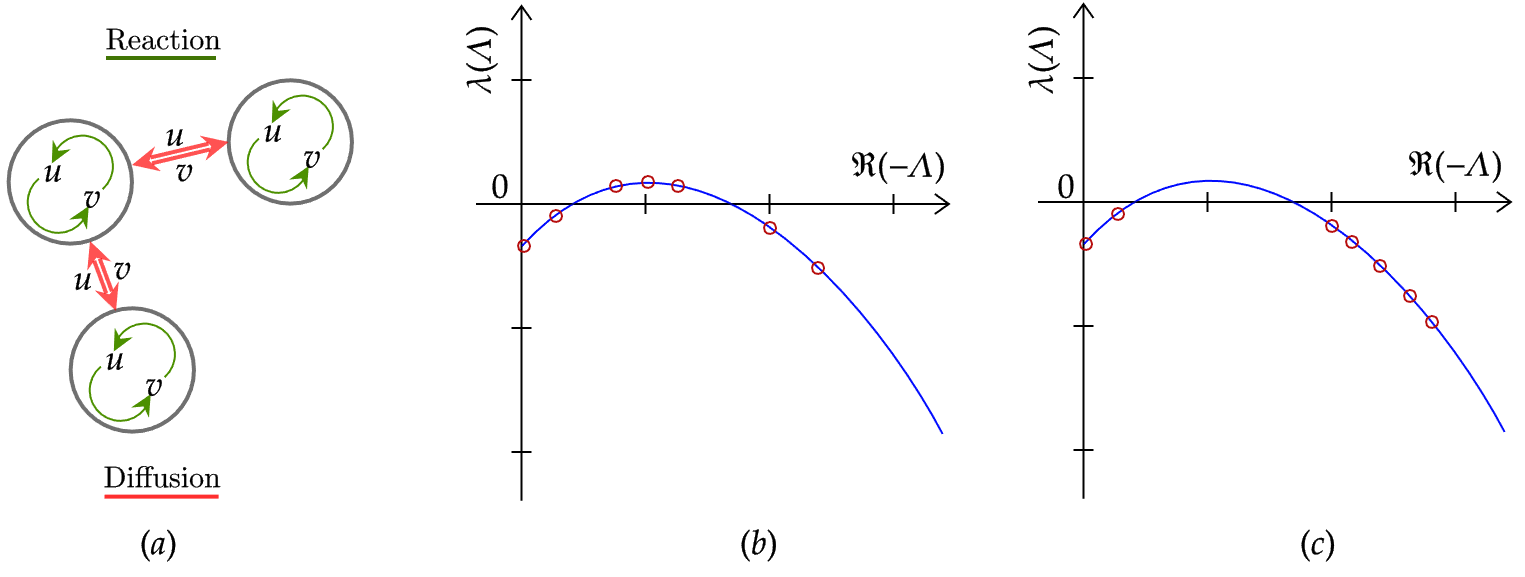}
\caption{Panel (a) depicts the reaction-diffusion network setting: the two species $u$ and $v$ interact within the nodes, which are well-stirred, while diffusing through the links. In panels (b-c) we visualize the conditions for the onset of the Turing instability: on networks, these are only necessary but not sufficient. In fact, the continuous dispersion relation can always be positive (blue line), while the discrete version (red circles) can be either positive (b) or negative(c).}
\label{fig:1}
\end{figure}

\begin{figure}
\centering
\includegraphics[width=\textwidth]{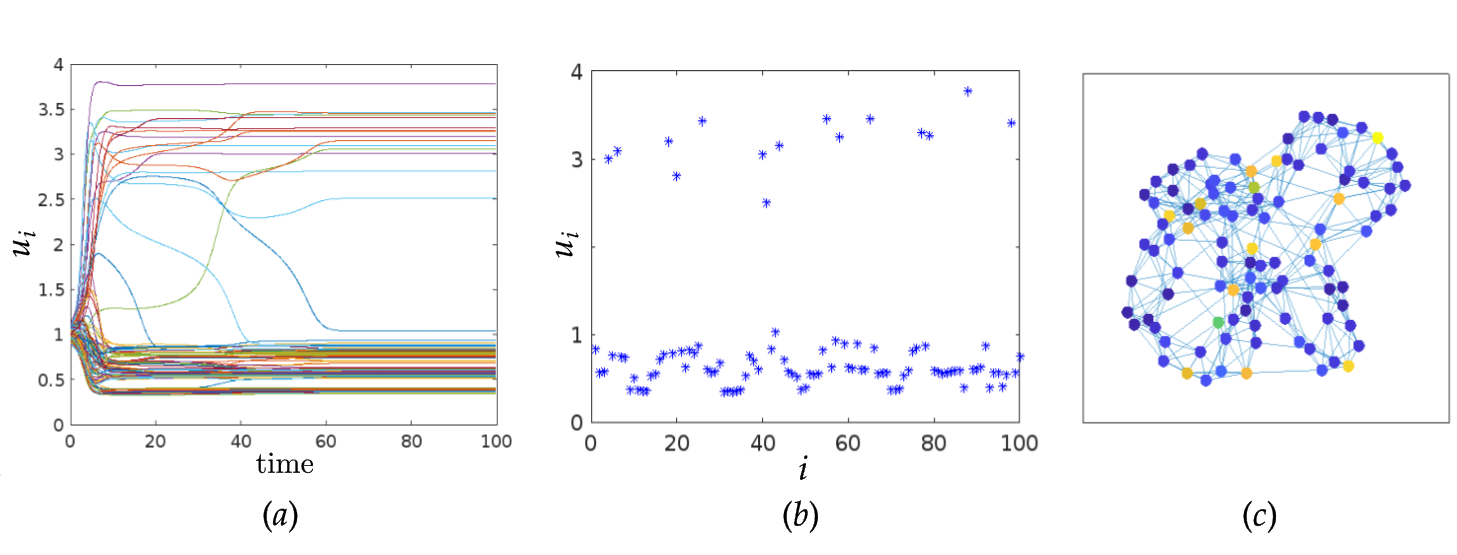}
\caption{Turing patterns for the Brusselator model \cite{PrigogineNicolis1967} on a Small-World network \cite{watts1998collective} of $100$ nodes. {   In panel (a), the time evolution of the concentration of the activator $u$ on each node is shown (qualitatively analogous results are obtained for the inhibitor $v$). The perturbation superimposed to the homogeneous equilibrium triggers the system unstable and the successive evolution makes an inhomogeneous state to spontaneously emerge. In panel (b), the final pattern, again for the concentration of species $u$ on each node, is displayed. The nodes are ordered as in the $1$D lattice from which the Small-World network is generated. It results into the segregation of nodes that are rich/poor of the activator $u$ (similarly for $v$). Finally, panel (c) depicts the concentration of the activator $u$ on the network nodes.}}
\label{fig:0}
\end{figure}

\begin{equation}
\label{eq:lin}
\begin{cases} 
\displaystyle \delta \dot{u}_i=f_u\delta u_i+f_v\delta v_i+D_u\sum_{j=1}^n L_{ij}\delta u_j,\\ 
\displaystyle \delta\dot{ v}_i=g_u\delta u_i+g_v\delta v_i+D_v\sum_{j=1}^n L_{ij}\delta v_j, \end{cases} \forall i=1,2,\dots,n,
\end{equation}
where $\delta u_i=u_i-u^*$, $\delta v_i=v_i-v^*$ are the inhomogeneous perturbations, i.e., node dependent, and $f_u$ stands for the derivative of the function $f$ with respect to variable $u$ computed at the fixed point (analogously for the others). By defining $\vec{\zeta}=(\delta u_1,\dots,\delta u_n,\delta v_1,\dots, \delta v_n)^\top$, we can rewrite the {  above equation} in compact form and obtain
\begin{equation}
\label{eq:2n2n}
\dot{\vec{\zeta}}=\left(\mathbf{J}_f \otimes \mathbb{I}_n  +\mathbb{D}\otimes \mathbf{L} \right) \vec{\zeta} \, , 
\end{equation}
where $\mathbb{D}=\left(
\begin{smallmatrix}
 D_u & 0\\
 0 & D_v
\end{smallmatrix}
\right)$, $\mathbf{J}_f=\left(
\begin{smallmatrix}
 f_u & f_v\\
 g_u & g_v
\end{smallmatrix}
\right)$ is the Jacobian matrix of the isolated system, $\mathbb{I}_n$ is the $n \times n$ identity matrix and $\otimes$ is the Kronecker product. Equation \eqref{eq:2n2n} determines the evolution of the perturbation: the system will relax back to the homogeneous equilibrium or exponentially escape toward another attractor, depending on whether the largest real part of the spectrum of the $2n\times 2n$ matrix $\mathbf{J}_f\otimes\mathbb{I}_n + \mathbb{D}\otimes\mathbf{L}$ is positive or negative. However, the basis on which we have written the above matrix is, {in general, not suitable} to study the evolution of the perturbation, as the (potentially) unstable modes are not decoupled. It turns out that there is a natural basis to handle the above linear problem. This is the basis formed by the eigenvectors of the Laplacian, in analogy with what is customarily done for a continuous support where the Fourier basis is routinely employed to ``diagonalize'' the Laplacian operator. In this way, the $2n\times 2n$ system is decomposed into $n$ systems of dimension $2$. For $\alpha=1,...,n$, we then obtain
\begin{equation}
\dot{\vec{\xi}}_\alpha=\left[\mathbf{J}_f+\Lambda^{(\alpha)}\mathbb{D}\right] \vec{\xi}_\alpha=:\mathbf{J}_\alpha\vec{\xi}_\alpha \, ,
\label{eq:new_pert1}
\end{equation}
where $\vec{\xi}_\alpha=(\hat{ u}_\alpha,\hat{ v}_\alpha)^\top$ is the perturbation {expressed} in the new basis and $\Lambda^{(\alpha)}$ is the $\alpha$-th eigenvalue of $\mathbf{L}$, which is either zero or negative, {associated to the eigenvector $\phi^{(\alpha)}$}. Let us remark that $\Lambda^{(1)}=0$ is the eigenvalue with the largest real part, associated to  the homogeneous eigenvector $\vec{\phi}^{(1)}\propto (1,\dots,1)^\top$; hence, the projection of the perturbation along this direction describes the dynamics of the uncoupled system. By analyzing the stability of matrix $\mathbf{J}_{\alpha}$, one can find the conditions for the emergence of patterns. Indeed, the maximum real part of the spectrum of $\mathbf{J}_{\alpha}$ as a function of $\Lambda^{(\alpha)}$ defines the {\em dispersion relation}, $\lambda = \max_\alpha  {\Re}\sigma(\mathbf{J}_{\alpha})$, where we denoted the set of eigenvalues by $\sigma(\mathbf{J}_{\alpha})$. Note that, the support being discrete, {the Laplacian spectrum will also be discrete, hence} there {can be} finite size effects: {  both dispersion relations (continuous and discrete) could be positive,
 thus implying the existence of a Turing instability (see Fig. \ref{fig:1}(b)); however, the discrete dispersion relation could be negative due to its intrinsic finiteness (see Fig. \ref{fig:1}(c))}, while the continuum counterpart proves positive. A thorough analysis of the finite size effects on the Turing instability regions has been carried out by Ide et al. in \cite{ide2016turing}. 
The strength of Eq.~\eqref{eq:new_pert1} relies on the fact that the whole information about the support, i.e., the network, is captured by the {eigenvalues of the Laplacian.} 

Also for networked systems, the perturbation is expanded on the basis of the Laplacian and its eigenvectors will affect the final shape of the pattern, exactly as the critical Fourier mode affects the pattern on a continuous support{  ; an example of such a pattern is shown in Fig. \ref{fig:0}.}

In the next Section, we will give an overview of what has been done in the field over the last 15 years, after the work of Nakao and Mikhailov \cite{NM2010}{  , focusing} our attention on three distinct settings which, we believe, allow one to fully appreciate the power of dealing with  networked supports.

\section{Ductility of the network approach: {  directed, multi-layer and time-varying topologies}}\label{sec:extensions_network}

As discussed at the beginning of Sec. \ref{sec:1st_approach}, already Turing's original paper \cite{Turing} contained the seeds of the network approach {   to the instability. This seminal contribution} inspired {   several other} works \cite{OS1971,OthmerScriven74,goras1995a,goras1995b,goras1995c,horsthemke2004,horsthemke2004b} {   where the discrete nature of the hosting support became progressively more evident}. {   The first extensions \cite{NM2010,pastorsatorrasvespignani2010} of Turing's ideas to complex networks - finally depicted under a modern perspective  -   triggered an intense activity which led to a vast load of works}. The aim of this section is to present those recent advances. There are three settings, namely, directed, multi-layer and temporal networks, that we think are particularly relevant to explain the power of the network framework {   when applied in conjunction with Turing's recipe}. First, because the continuous-discrete dichotomy {   to which we alluded above is now, at least in part, }  broken, as {   in most relevant settings complex networks do not admit} a trivial continuous embedding; then,  {   networks are always invoked to model}  complex systems {   subject to non-reciprocal interactions}  \cite{newmanbook2}. Multi-layer \cite{BianconiMultilayer} and time-varying \cite{holme_temporal} {   represent additional directions of exploitations beyond the original scenario envisaged by Turing}. In this section, we will mainly focus on these {   latter aspects. At the end, we will also provide} a general overview of other important {   lines of research that have emerged as an obvious byproduct of the extension of the Turing paradigm to network science}. 

\subsection{Directed and non-normal topologies }\label{sec:dir_nn}

When the network is directed, the Laplacian matrix is asymmetric and thus its spectrum is, in general, complex. Let us first assume that the Laplacian is diagonalizable, namely, there exist complex vectors $\vec{\phi}^{(\alpha)}$ associated to eigenvalues $\Lambda^{(\alpha)}=\Lambda^{(\alpha)}_{\Re}+i\Lambda^{(\alpha)}_{\Im}$, for $\alpha=1,\dots, n$. Let us remark that $\Lambda^{(1)}=0$ is still the eigenvalue with the largest real part and, being the eigenvector $\vec{\phi}^{(1)}\propto (1,\dots,1)^\top$, the projection of the perturbation along this direction describes once again the dynamics of the uncoupled system. With those assumptions, the linear stability analysis of the previous Section still stands, with the difference that now Eq. \eqref{eq:new_pert1} becomes 
\begin{equation}
\dot{\vec{\xi}}_\alpha=\left[\mathbf{J}_f+\Big(\Lambda^{(\alpha)}_{\Re}+i\Lambda^{(\alpha)}_{\Im}\Big)\mathbb{D}\right] \vec{\xi}_\alpha=\mathbf{J}_\alpha\vec{\xi}_\alpha .
\label{eq:new_pert1_new}
\end{equation}

{  It can be shown that the imaginary part of the Laplacian spectrum has the effect of increasing the maximum of the real part of the eigenvalue of $\mathbf{J}_{\alpha}$, i.e., the dispersion relation~\cite{Asllani1}. It is possible to find an analytical condition for the instability, which is summarized by the following compact inequality:  \begin{equation}
    \label{eq:instregion}S_2(\Lambda^{(\alpha)}_{\Re})[\Lambda^{(\alpha)}_{\Im}]^2 < -S_1(\Lambda^{(\alpha)}_{\Re}),
\end{equation} where $S_1$ and $S_2$ are polynomials in $\Lambda^{(\alpha)}_{\Re}$ of degree $4$ and $2$, respectively, whose coefficients depend on the model parameters, i.e., the entries of $\mathbf{J}_f$, and the diffusion coefficients (see~\cite{Asllani1} for the details of the derivation and for the explicit list of coefficients). This allows to obtain an alternative representation of the instability onset in the complex plane, see Fig.~\ref{fig:nat_comm}(a), where the role of the imaginary part of the Laplacian spectrum can be fully appreciated. The consequence of the analysis is that it is easier to generate Turing patterns on directed discrete support than on their symmetric analogues. In Fig.~\ref{fig:nat_comm}(b), we depict the effect of the asymmetric topology on the dispersion relation (red circles), which lies above its continuous counterpart (blue curve). {    This is a general conclusion: any undirected network yields a dispersion relation which lies on top of its continuous analogue. At variance, the dispersion relation obtained for a Laplacian spectrum with non-zero imaginary part detaches from the corresponding curve obtained for the reference symmetric case.}\footnote{  Let us point out that not all directed networks displays a complex spectrum, but directed networks with real spectra are non-normal and non-normality is responsible for another intriguing effect to which we shall make reference in the following.} {  Another consequence of dealing with an asymmetric topology is that} the imaginary part of the ensuing Laplacian spectrum may yield oscillatory Turing patterns with only two species, a phenomenon that cannot take place on a symmetric support, as discussed in the Introduction.} {   An example of oscillatory pattern obtained under these operating conditions is shown in the inset of Fig. \ref{fig:nat_comm}(b). This path to the instability, termed \textit{topology-driven}, has been first observed by Asllani et al. in \cite{Asllani1} and it is analogous to what was found on a continuous support when the system is subject to an external drift, as reported by Rovinsky and Menzinger back in the 90s \cite{rovinsky1992chemical}. Very recently, He and Su \cite{he2024spatiotemporal} considered the setting of symmetric networks with advection (a scenario which is in practice analogous to a directed network) obtaining results in line with those reported by Asllani et al. \cite{Asllani1}. Interestingly, the work by Fruchart et al. \cite{fruchart2021non} on phase transitions represents a further confirmation of the conclusions \cite{Asllani1}. Indeed the claim that directed networks increase the formation of patterns has been further corroborated in \cite{Asllani2,ritchie2023turing,carletti}.} 

\begin{figure}[h!]
\centering
\includegraphics[width=\textwidth]{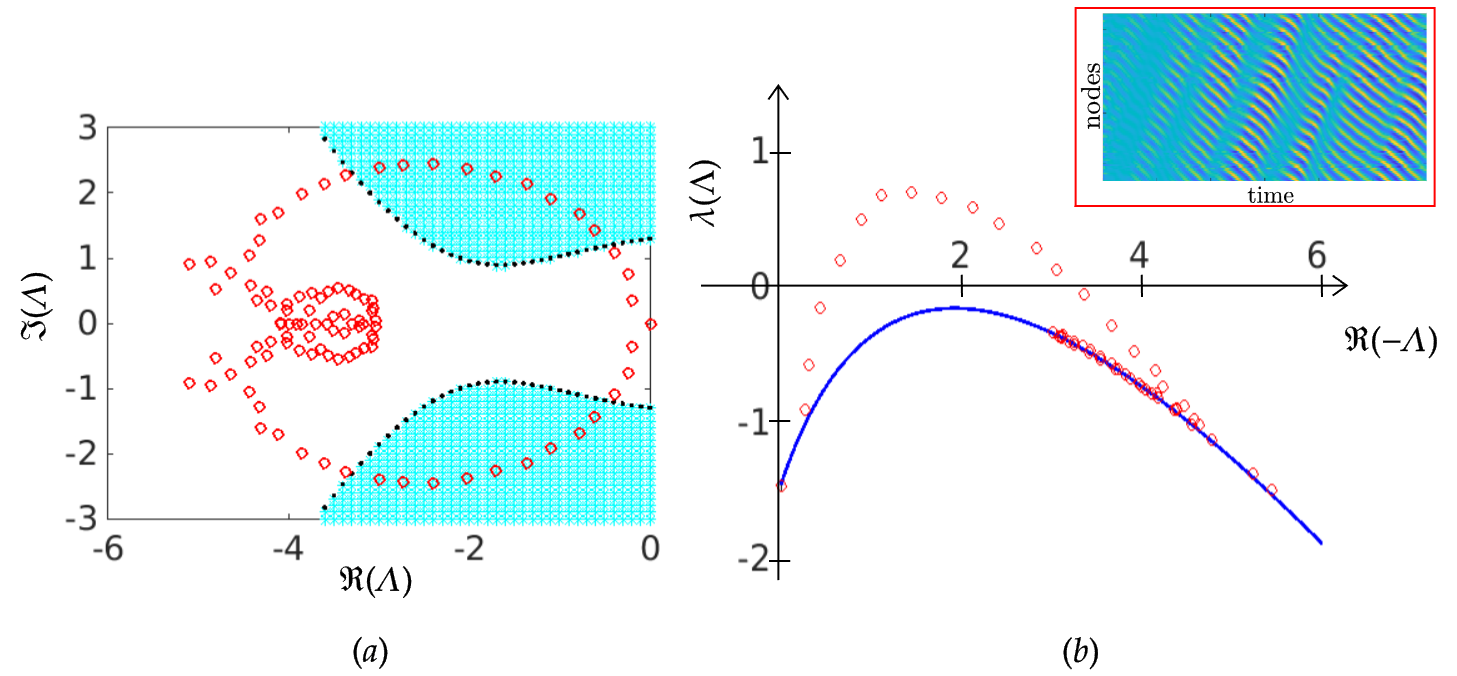}
\caption{{   Panel (a) shows the instability region in the complex plane. When condition \eqref{eq:instregion} is satisfied, the spectrum of the Laplacian (red circles) reaches the instability region (cyan region). Let us observe that in this case the instability region does not intersect the negative real axis and, thus, a symmetric network would not achieve instability, the spectrum being real. The same setting is represented through the real dispersion relation, shown in panel (b). Here, we can observe that the effect of the non-zero imaginary part of the Laplacian spectrum consists in a detachment from the symmetric case (blue curve), allowing the dispersion relation (red circles) to reach the instability region. The inset of panel (b) shows an example of an oscillatory Turing pattern for the Brusselator model \cite{PrigogineNicolis1967} on a directed Newman-Watts network \cite{newman1999scaling} of $100$ nodes. The oscillations are possible because of the directed topology, as explained in the text. Inspired by Fig. 1 of \cite{Asllani1}.}}
\label{fig:nat_comm}
\end{figure}

Let us now discuss an additional effect that an asymmetric topology has on the dynamics of Turing patterns. When a matrix is asymmetric, it is in general non-normal, i.e., $A^\top A\neq AA^\top$, where $A^\top$ denotes the transpose of $A$. The above relation implies that $A$ is non-diagonalizable through an orthonormal transformation, i.e., it is either diagonalizable but without an orthogonal basis or non-diagonalizable \textit{tout court} \cite{trefethen}. Such property can have important effects on the dynamics: in fact, as first found by Trefethen et al. in the context of fluid dynamics \cite{trefethen2}, a solution predicted stable by a linear stability analysis can, instead, go unstable after a finite perturbation. The cause of such effect is a {  transient growth, that non-normality can induce even in the linear regime. To better understand this phenomenon, consider the simple $2\times 2$ linear system $\dot{\vec{x}}=A\vec{x}$, where $A$ is a real stable matrix, i.e., with a negative spectrum. Let us introduce the \textit{numerical abscissa} $\omega(A)=\max(\sigma((A+A^\top)\slash 2))$, i.e., the maximum eigenvalue of its symmetric part, which controls the short-term behavior of the system \cite{top_resilience}. Given that $A$ is stable, $\omega$ can be positive only if $A$ is non-normal. Indeed, such transient growth is caused by a positive numerical abscissa: after the equilibrium is perturbed, the solution grows at short times, but then relaxes back to the initial state, since the system is 
linear and stable by assumption. Such an effect can be visualized in Fig. \ref{fig:2bis}(a). This scenario holds in general} and it has been observed also in ecology \cite{neub_cas}, neuroscience \cite{murphy2009balanced,hennequin2014optimal}, quantum physics \cite{ferraz2023revisiting} and engineering \cite{ganguli2008memory,baggio2020efficient,lindmark2020centrality}, to name a few. 
In the context of Turing patterns, it was proven by Neubert et al. \cite{murray_neub_cas} that a positive numerical abscissa for $\mathbf{J}_f$ (which, let us recall, is a stable matrix) is a necessary condition for Turing instability. Klika discussed the importance of the transient on the Turing mechanism on continuous support \cite{klika2017significance}, while Biancalani et al. observed the amplification of the transient due to noise \cite{biancalani2017giant}. Then, Asllani and Carletti showed that such a transient growth effect is further enhanced when the reaction-diffusion system is defined on top of non-normal networks \cite{top_resilience}, which ultimately increases the region in which Turing patterns emerge \cite{jtb}. In fact, when the homogeneous state is perturbed, the transient growth is enhanced by the non-normal network topology and it may result in a pattern, even when a linear stability analysis would predict no pattern formation, as depicted in Fig. \ref{fig:2bis}(b). Let us observe that this phenomenon is not {  a proper} diffusion-driven instability, even if the resulting pattern resembles a Turing one{  . Indeed, the homogeneous state remains stable also in presence of the diffusive terms (i.e., negative dispersion relations) but its stability basin is very small; hence it can happen that a relatively small perturbation is amplified driving thus the emergence of an inhomogeneous solution resulting from the interplay between the} nonlinear dynamics and the network non-normality. Such effect is even greater in the presence of noise, as shown by Nicoletti et al. \cite{duccio_stoch}, because the solution, which would naturally move toward the stable equilibrium, is continuously ``kicked'' away from the former by the noise. 

Let us conclude this part by relaxing {   the assumption on the existence of an eigenbasis for the Laplacian matrix. One can indeed show~\cite{dorchain2023pattern} that when the network is degenerate, namely the Laplacian matrix admits Jordan blocks with repeated eigenvalues, a linear stability analysis can be still performed to yield a dispersion relation from the set of existing eigenvectors}. Observe that a similar idea has been used by Nishikawa and Motter \cite{mott_nish2} to study the synchronization of coupled nonlinear systems on directed networks. Moreover, the use of generalized eigenvectors can help in pattern reconstruction, as found by Dorchain et al. \cite{dorchain2023pattern}. All this makes the setting particularly relevant for applications, given that, as found by Asllani et al. \cite{asllani2018structure}, real-world networks are non-normal and, often, degenerate. Over the last few years, a great effort has been dedicated to understanding how to measure the directedness and the non-normality of a network \cite{mackay2020directed,pilgrim2020organisational} and which features of a non-normal network are most relevant for the dynamics \cite{nartallo2024broken,duan2022network}.

\begin{figure}
\centering
\includegraphics[width=\textwidth]{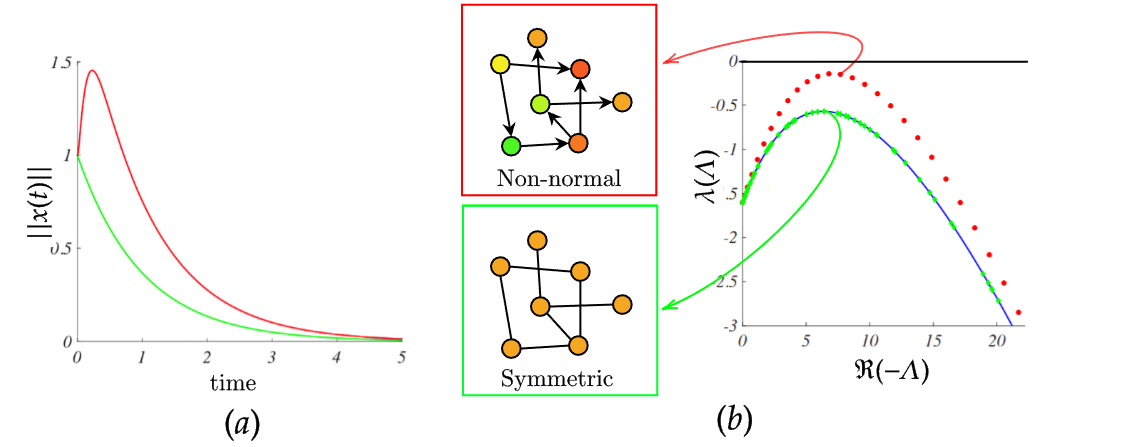}
\caption{{   In panel (a), we show the effect of non-normality on a linear dynamics, namely on the $2\times 2$ linear system $\dot{\vec{x}}=A_i\vec{x}$, with $i={1,2}$. Both cases involve stable matrices, $A_1$ and $A_2$, and the two solutions relax to the equilibrium after being perturbed. Nonetheless, $A_1$ is such that its numerical abscissa is positive, i.e., $\omega(A_1)>0$. Then, the system undergoes a transient growth (red curve), at variance with the evolution observed for $A_2$, whose numerical abscissa is negative (green curve). When the dynamics is nonlinear, such effect can dramatically change the fate of the system and, in panel (b), we depict a case study relevant to Turing mechanism. Here, the linear stability analysis would predict no patterns for both systems, as both cases (symmetric - green circles, directed - red circles) return negative values for the dispersion relation. The directed network is, however, non-normal and the transient dynamics caused by non-normality allows the system to "escape" from the basin of attraction of the homogeneous solution, yielding asymptotic Turing patterns (shown in the network). Adapted from Figs. 2 and 4 of \cite{jtb}, where the model is the Brusselator \cite{PrigogineNicolis1967}; reproduced with permission.}}
\label{fig:2bis}
\end{figure}

\subsection{Beyond networks: {  multiplex and multi-graphs}}\label{sec:mult}

\begin{figure}[h]
\centering
\includegraphics[width=\textwidth]{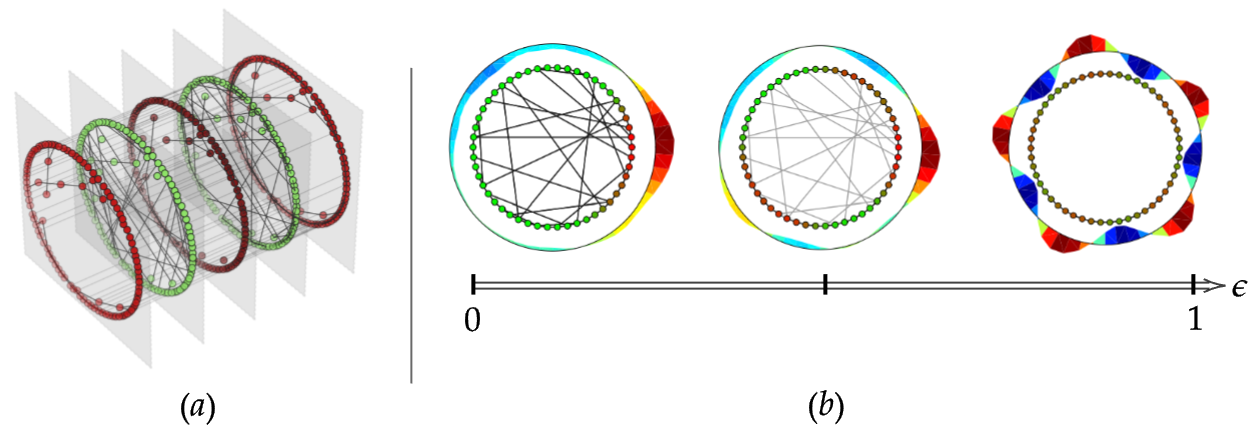}
\caption{  
{   Panel (a), illustrative example of a layer-homogeneous fixed point as obtained for the case of a multiplex network composed of five adjacent Watt-Strogatz layers. Here the Brusselator model is assumed to govern the reactive dynamics. Remarkably enough, the system spontaneously evolves towards an asymptotic state where the depicted species displays a different density (as exemplified by the color associated to individual nodes) depending on the layer it belongs to.  Panel (b) reports on the emergence of localized patterns for the Brusselator model evolving on a multi-graph. More specifically, we assume that species diffuse on a multigraph network, namely the activator has access to the links encoded by the adjacency matrix $A_0$, while the inhibitor moves by using the links associated to the network described by the matrix $A(\epsilon) = A_0 + \epsilon(A_1 - A_0)$, where $\epsilon \in [0,1]$ gauges the relative importance of the two supports. Stated differently, we can tune the channels available to the inhibitor to coincide with those used by the activator by setting $\epsilon=0$, or to allow for a completely different path of movement if $\epsilon=1$. The ensuing patterns can be modulated, from localized to periodic depending on the choice of $\epsilon$. The obtained patterns (color intensity displayed at the node level) are anticipated by looking at the characteristics of the most unstable eigenvector as revealed by a linear stability analysis. For the three inspected examples, the relevant eigenvector is plotted with an apposite color code. It also encloses the network setting to which it refers to (plot surrounding) to facilitate the comparison with the ensuing dynamical pattern. Panel (a) is Fig. 1 of \cite{busiello_homogLturing}, while panel (b) is Fig.4 of \cite{Asllani2016}; reproduced with permission.} 
}
\label{fig:multi}
\end{figure}

In some real-world applications, the system cannot be encoded by a network{  , which may not be general enough} to fully take into account the underlying complexity, see e.g., the transportation systems~\cite{kurant2006layered,zou2010topological}, the learning organization in the brain~\cite{bullmore2009complex}, the economic market~\cite{YANG20092435} or the emergent dynamics in social communities~\cite{social}.  Those examples all share a common feature, namely, the system can be thought as composed of several ``layers'', each one  {possibly carrying different information about the system: for example, in a transportation network each layer can encode for the different kinds of transportation means used by agents to relocate among locations, i.e., network nodes. It can then happen that the same node appears in different layers: think of a bus station connected to a train station and to an underground one. The presence of those homologous nodes allows agents to pass from one layer into the others}. Scholars have thus defined and studied multiplex networks, {   which are made of $m>2$ networks (i.e., the layers) composed by the same nodes, each layer having its own connectivity structure (i.e, \textit{intra-layer links}). Because each node is ``repeated'' in all the layers, the \textit{inter-layer} connectivity is trivial: all the links connect the same node over different layers}~\cite{CozzoEtal,BianconiMultilayer,boccaletti2014structure,battiston2014structural}. Also on continuous support, there are studies of pattern formation on multi-layer geometry, e.g., \cite{krause2020turing,duda2023modelling,diez2024turing}, which could be mapped to a multiplex network.
Diffusion processes on multiplex networks depend thus on the diffusion coefficients in each layer but also on those responsible for the movement across layers. Gómez et al. \cite{gomez2013diffusion} have studied such a process by defining the supra-Laplacian matrix and shown that, for some configurations of the layer topology and inter-layer diffusion coefficient, one can face super-diffusion, i.e., the spreading process on the multiplex is faster than the one on each layer acting separately. The impact of inter-layer diffusion on the emergence of Turing patterns has been studied in reaction-diffusion systems defined on multiplex networks~\cite{Asllani2014}.

{  
Let us assume, for simplicity, to deal with two layers. Then Eq.~\eqref{eq:react_diff_eq} should be modified to account for the possibility of the species to diffuse across contiguous layers. This is accomplished as follows:
\begin{equation}
\begin{cases}
\dot{u}^K_i&= f(u^K_i,v^K_i) + D_u^K\displaystyle\sum_{j=1}^{n}L_{ij}^K u^K_j + D_u^{12} \left(u^{K+1}_i-u^K_i\right), \\
\dot{v}^K_i&= g(u^K_i,v^K_i) + D_v^K\displaystyle\sum_{j=1}^{n}L_{ij}^K v^K_j + D_v^{12} \left(v^{K+1}_i-v^K_i\right),
\end{cases}\, 
\label{eq:reac_diff}
\end{equation}
where $u_i^K$ (resp. $v_i^K$) denotes the density of species $u$ (resp. $v$) in the $i$-th node on the $K$-layer, with $K=1,2$ and setting $K+1$ to be $1$ for $K=2$. We thus assume that reactions, encoded by nonlinear functions $f(u^K_i,v^K_i)$ and $g(u^K_i,v^K_i)$, solely take place between individuals that happen to share the same node $i$ and layer $K$, $D_u^K$ (resp. $D_v^K$) rules the diffusion of species $u$ (resp. $v$) across nodes of the $K$-layer, while $D^{12}_u$ (resp. $D^{12}_v$) the diffusion of species $u$ (resp. $v$) across homologous nodes belonging to different layers. Hence $L_{ij}^K=A_{ij}^K-k^K_i\delta_{ij}$ stands for the Laplacian matrix on the layer $K$. If the inter-layer diffusion is silenced, which implies setting $D_u^{12}=D_v^{12}=0$, the layers are decoupled and we obtain again system~\eqref{eq:react_diff_eq}.
As before, we can perform a linear expansion of the former equations close to the homogeneous equilibrium $(u_K^*,v_K^*)$, to obtain
\begin{equation}
\frac{d}{dt}\left( \begin{array}{ccc}{\delta\boldsymbol{u}}\\{\delta\boldsymbol{v}}
 \end{array} \right)=\left( \begin{array}{ccc}
f_u \mathbf{I}_{2n} + \boldsymbol{\mathcal{L}}_u +D_u^{12}\boldsymbol{\mathcal{I}} & f_v \mathbf{I}_{2n}\\
g_u \mathbf{I}_{2n} & g_v \mathbf{I}_{2n} + \boldsymbol{\mathcal{L}}_v+D_v^{12}\boldsymbol{\mathcal{I}}
 \end{array} \right)\left( \begin{array}{ccc}
\delta\boldsymbol{u}\\\delta\boldsymbol{v}
 \end{array} \right),
 \label{eq:lin_prob2}
\end{equation}
where $\delta \boldsymbol{u} = (\delta u_1^1,\dots, \delta u_n^1,\delta u_1^2,\dots, \delta u_n^2)^\top$, and similarly for $\delta \boldsymbol{v}$. Moreover, $\boldsymbol{\mathcal{I}}=\left(\begin{smallmatrix} -\mathbb{I}_{n} & \mathbb{I}_{n}\\ \mathbb{I}_{n} & -\mathbb{I}_{n}\end{smallmatrix}\right)$ accounts for the diffusion across homologous nodes in layers $1$ and $2$, and the multiplex Laplacian for the species $u$, defined as
\begin{equation*}
\boldsymbol{\mathcal{L}}_u=\left( \begin{array}{ccc}
D_u^1\mathbf{L}^1 & \mathbf{0}\\
\mathbf{0} & D_u^2\mathbf{L}^2
 \end{array}\right)\,,
\end{equation*} 
accounts for the diffusion of species $u$ in each layer. A similar operator, $\boldsymbol{\mathcal{L}}_v$, is associated to species $v$. Let us observe that
$\boldsymbol{\mathcal{L}}_u +D_u^{12}\boldsymbol{\mathcal{I}}$ is the supra-Laplacian introduced in~\cite{gomez2013diffusion}.

The emergence of Turing pattern can be proven by studying the linear system~\eqref{eq:lin_prob2}. This ultimately relies on the computation of the spectrum of the involved matrix. Since Laplacian operators referred to different layers do not (generically) commute, they cannot be diagonalized by resorting to the same eigenbasis. Hence, the usual strategy of projecting the perturbations on a suitable basis cannot be here adopted. There are, however, relevant cases where analytical progress is possible, notably when the inter-layer diffusion is much smaller than the intra-layer one. A perturbative analysis can be performed which enables one to compute the spectrum of the whole system as a function of the spectra of the operators associated to each individual layers and on the coefficients $D^{12}_u$ and $D^{12}_v$. Inter-layer diffusion can enhance the emergence of collective patterns, otherwise absent in the case of uncoupled layers. Let us notice that patterns can also vanish due to the couplings among distinct layers. The interested reader can refer to~\cite{Asllani2014,busiello_homogLturing} for a more detailed description of the relevant mathematical issues and to Fig. \ref{fig:multi}(a) for a graphical representation of the results. 

Let us observe that a similar idea can be used to tackle the problem of pattern emergence on multi-graphs, i.e., where each species uses a different set of connections to move across the nodes where reactions occur. Turing patterns on multi-graphs have been studied by Kouvaris et al. \cite{KHDG}, Asllani et al.~\cite{Asllani2016} and, several years later, by Lei and He \cite{lei2021patterns}. This framework ultimately relies on replacing the single Laplacian matrix in Eq.~\eqref{eq:react_diff_eq} with two Laplacian matrices, each ruling the diffusion process of a given species (see Fig.\ref{fig:multi}(b) for a schematic visualization of the reference setting and \cite{Asllani2016} for a complete account of the underlying theory).}

In conclusion, and building on several complementary observations, it has been argued that the hierarchical organization of the space where dynamics occurs plays a role of paramount importance in seeding the patterns observed in nature. This has been also recently claimed by Song et al. for applications of interest to ecology \cite{song2023cross}.

\subsection{Temporal networks}

So far we have considered reaction-diffusion processes defined on static networks, namely, neither the links nor the number of nodes {   change} over time. However, in some cases, the networks evolve in time and one has to consider time-varying networks, where links are created, destroyed, rewired or their weights can depend on time \cite{holme_temporal,masuda2016guide}. Such structures find many applications in nonlinear dynamics, including synchronization \cite{ghosh2022synchronized} and control \cite{li2017fundamental}. Let us stress that the varying number of nodes can be simply handled by considering the existence of a (very) large reservoir of nodes, being part of the system but disconnected as long as they do not interact with the main core of the network. Hence, the creation of a new node is actually the connection of an existing node with the core and the destruction of a node is the disconnection of the latter. {   Ultimately, this amounts to modify Eq.~\eqref{eq:react_diff_eq} by explicitly accounting for the time varying nature of the underlying support to eventually obtain
\begin{equation}
\label{eq:reac_diffacc}
    \begin{cases}
\dot{u}_i(t)&=  f(u_i,v_i) + D_u\displaystyle\sum_{j=1}^{N}{L}_{ij}(t) u_j, \\
\dot{v}_i(t)&= g(u_i,v_i) + D_v\displaystyle\sum_{j=1}^{N}L_{ij}(t) v_j.
    \end{cases}\, 
\end{equation}
}

The static network assumption is surely valid whenever the network evolution is very slow with respect to the natural time scale of the dynamical system running on top of the network. In the remaining cases, one has to consider the proper time evolution of the network and its interaction with the reaction-diffusion process. Scholars have also considered the opposite limit, where the network evolution is very fast. In particular, by using fast switching networks{  , i.e., at given time points the network instantaneously passes from one static configuration to a different one,} synchronization dynamics has been studied \cite{stilwell2006sufficient}, but also the emergence of Turing patterns \cite{PABFC2017} and oscillation death \cite{LFCP2018}. {   To make this assumption explicit, one has  to modify~\eqref{eq:reac_diffacc} by introducing a second time scale, by means of a scaling parameter $\epsilon>0$, as given in the following 
\begin{equation}
\label{eq:reac_diffaccfast}
    \begin{cases}
\dot{u}_i(t)&=  f(u_i,v_i) + D_u\displaystyle\sum_{j=1}^{N}{L}_{ij}(t/\epsilon) u_j,\\\\
\dot{v}_i(t)&= g(u_i,v_i) + D_v\displaystyle\sum_{j=1}^{N}L_{ij}(t/\epsilon) v_j.
    \end{cases}\, 
\end{equation}
}

These works rely on the application of the averaging theorem for dynamical systems, which roughly states that if one can prove the existence of some limiting ``average system''  {(in this case an average adjacency matrix $\langle A \rangle$ to which one can associate a Laplacian matrix)}  and if the frequency of the time-varying one is large enough  {(i.e., $\epsilon \ll 1$)}, then the two systems exhibit a similar behavior where precise bounds on the orbits difference can be provided. Stated differently, if the average system does exhibit Turing patterns, i.e., any (sufficiently small) heterogeneous perturbation about the homogeneous fixed point is amplified, then the time-varying one does the same and the homogeneous fixed point is again unstable  {(see Fig.~\ref{fig:Fig1TPTimeVarying})}. Let us observe that in \cite{chavez2005weighted,amritkar2006synchronized,zhang2021designing}, authors have considered the case of commuting temporal networks. More precisely, they assumed that the time-varying Laplacian matrices {  commute for every time $t$}. This allows one to relax the assumption of the fast network connectivity at the price of introducing a somehow restrictive working hypothesis, which in turn amounts to state that Laplacian eigenvectors do not  evolve over time. Namely, there exists a fixed eigenbasis and only the eigenvalues change in time. A similar result has been obtained by Van Gorder in \cite{vangorder2}. 

\begin{figure}[ht!]
\begin{center}
\includegraphics[width=\textwidth]{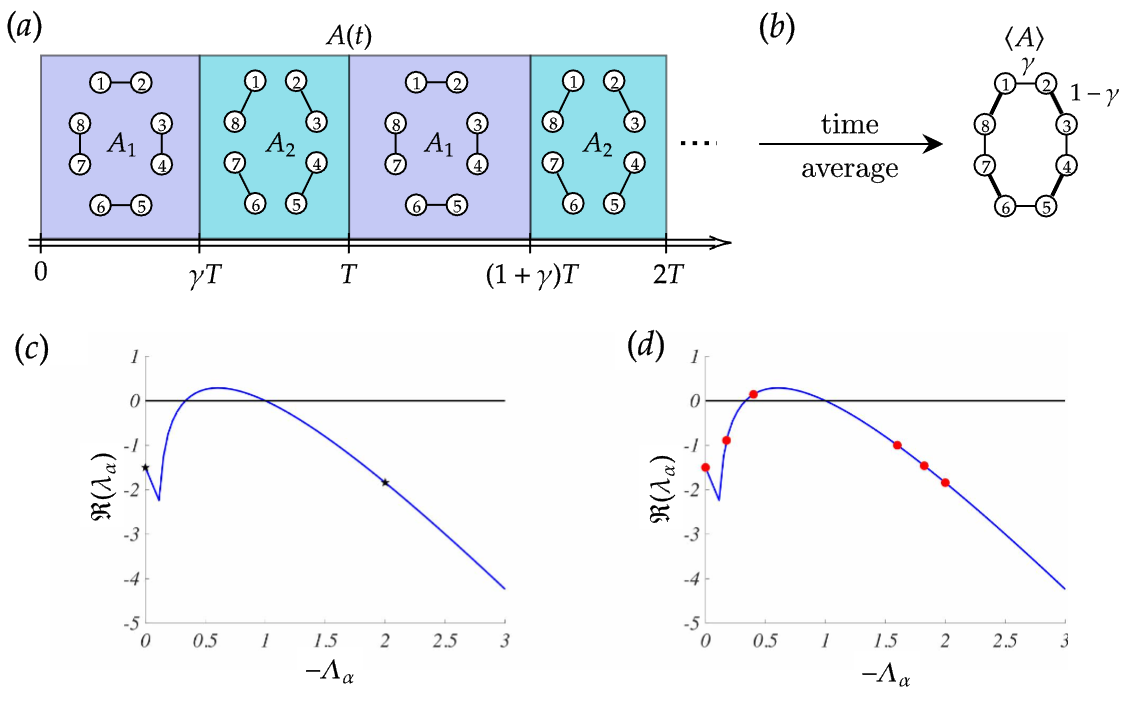}
\end{center}
\caption{{  Time-varying network. Panel a): $T$-periodic network built from two static networks with adjacency matrices $\mathbf{A}_1$ and $\mathbf{A}_2$ encompassing for $8$ nodes each. In the network encoded by $\mathbf{A}_1$ (purple), symmetric edges are drawn between the pairs $(1,2)$, $(3,4)$, $(5,6)$ and $(7,8)$. The second network,  embodied in matrix $\mathbf{A}_2$ (light blue), links nodes $(8,1)$, $(2,3)$, $(4,5)$ and $(6,7)$.
For $t\in[0,\gamma T)$ the $T$-periodic network coincides with $\mathbf{A}_1$, $\mathbf{A}(t)=\mathbf{A}_1$, while in $[\gamma T,T)$ we set $\mathbf{A}(t)=\mathbf{A}_2$. The time varying network is then obtained by iterating the process in time. 
Panel b): the ensuing time average network $\langle \mathbf{A}\rangle =  \gamma \mathbf{A}_1+(1-\gamma)\mathbf{A}_2$, where link weights are a measure of the activation time of the link in each fixed network. Panel c): the dispersion relation ($\max\Re \lambda_{\alpha}$ vs. $-\Lambda^{\alpha}$) for the two networks $\mathbf{A}_1$ and $\mathbf{A}_2$ (black stars) and for the continuous support case (blue curve), we can observe that the dispersion relation is negative and thus on each fixed network, patterns can not develop. Panel d):  dispersion relation for the average network $\langle A \rangle$ (red circle) and for the continuous support case (blue curve). We can now appreciate the existence of an eigenvalue $-\Lambda_\alpha$ associated to a positive value of the dispersion relation; hence the average network supports Turing patterns. 
Inspired by Fig. 1 of \cite{PABFC2017}, where the model is the Brusselator \cite{PrigogineNicolis1967}.}}
\label{fig:Fig1TPTimeVarying}
\end{figure}

In a recent work~\cite{CARLETTI2022112180}, authors have considered the general case, where no assumptions are made on the relative pace of the network evolution and the natural time scale of the dynamical system. The key result is the generalization of the so-called \textit{Master Stability Function} (derived in Sec. \ref{sec:2nd_approach}), from which one can determine the stability {  properties} of the reference homogeneous solution. Such an approach returns the dispersion relation when the solution is a stationary point, the latter containing an extra term describing the time variation of the eigenvectors. The results reported in~\cite{CARLETTI2022112180} suggest that the size of the parameters domain associated with Turing patterns shrinks once the system evolves on a time-varying network.

\subsection{Further advances and applications of Turing theory on networks}

In this Section, we will give a general overview of the main works outside the framework of directed, multi-layer and temporal networks, without discussing {  them in} details and referring the interested reader to the Bibliography. 

The network support provides a natural framework for several applications, including ecology and neuroscience, which have been studied also in the Turing setting. For example, Nakao and Mikhailov's work relies on the Mimura-Murray model \cite{NM2010}, inspired {  by} ecology, and the Gierer-Menhardt model \cite{GiererMeinhardt} is used in \cite{guo2021turing}. Turing patterns on networks have been found for other models, for example, the Brusselator model \cite{PrigogineNicolis1967}, inspired {  by} chemistry, in \cite{ji2020turing}, or the Fitzhugh-Nagumo model \cite{fitzhugh,nagumo}, from neuroscience, in \cite{zheng2020turing}. A series of works extended settings previously studied on continuous support to the network framework, such as stochastic patterns \cite{asllani_noise,asllani2013linear,xiao2023effects}, hyperbolic reaction-diffusion systems \cite{jop_carletti}, delay differential equations \cite{petit2015delay,petit2016pattern,chang2019delay,tian2020delay}, reduced systems \cite{CarlettiNakao} and cross-diffusion \cite{zheng2017pattern,kuehn2024cross}. The network framework also allows us to obtain Turing-like patterns without diffusion \cite{CENCETTI2020109707,carletti}, where the interactions are modeled by a different Laplacian matrix, called consensus Laplacian. Moreover, the theory has been further generalized in the framework of graphons  \cite{bramburger2023pattern}, i.e., continuous limits of networks. 

Turing patterns have been studied on different network topologies, such as modular networks by Siebert et al. \cite{siebert}, geometric networks by van der Kolk et {  al.} \cite{van2023emergence}, cartesian product networks by Asllani et al. \cite{asllani2015turing} and non-homogeneous Cellular Neural Networks (CNN) by Goraş et al. \cite{goras2017turing}. In all the above cases, the network structure was crucial to obtain the patterns. Moreover, it has been shown that the patterns on the network settings are often localized \cite{nicolaides2016self}, which is caused by the structure of the eigenvectors in certain network topologies, such as the scale-free ones \cite{hata2017localization}. Localized patterns can also be understood by studying the differentiation of a single node, as shown in \cite{wolfrum2012turing,mccullen2016pattern}. The effects of the network topology on the pattern have been further highlighted in several works: Mimar et al. used a fine tuning between the diffusion coefficients of the species and the topology \cite{mimar2019turing}, then Hutt et al. showed the impact of structural changes in the network on the final pattern \cite{hutt2022predictable}, Chang et al. showed the relation between the average degree and the pattern \cite{chang2022qualitative}{  , and, more recently, Pranesh et al. analyzed the effects of clustering \cite{Pranesh2024effect}.} 
The network settings can also be generalized as a metaplex \cite{estrada2020metaplex}, a structure consisting of nodes, inside which another network is present, connected to each other through long-range links. Such structure has proven to be useful in modeling ecosystems \cite{yeakel2014synchronisation,gross2020modern} and diffusion-driven patterns have been found for $3$ species \cite{hata_dispersal} and in the general case\cite{brechtel,baron2020dispersal}. Still, in the framework of ecology, some works have been carried out with the prey-predator model \cite{fernandes2012turing,liu2020turing}, and also considering cross-diffusion \cite{ghorai2016turing}. 

The network support is well-suited also for applications in control and it has inspired several works in this direction. Such framework has been exploited for pinning-control by Buscarino and collaborators \cite{chua_Turing2,control_frasca}, feedback control by Hata et al. \cite{control_hata} and, more recently, optimal control by Gao, Chang and collaborators \cite{control_holme,liu2022optimal}. Moreover, it is possible to change the network structure while obtaining the same pattern by preserving some key features, as it has been shown in the symmetric case by Cencetti et al. \cite{cencetti2018pattern} and for directed networks by Nicoletti et al. \cite{nicoletti2020generating}. Similar principles have also been applied in methods to infer missing edges by Ali et al. \cite{ali2022inferring}. Lastly, let us mention that Turing mechanism has also been applied in the context of epidemics to model the spreading process on network topologies {  \cite{duan2019turing,chang2020cross,zheng2022pattern,zhou2023pattern,he2023turing,sun2024dynamics}} and there is a line of research involving parameter identification and estimation that is evolving very rapidly \cite{he2024parameter,chang2024time,matas2024unraveling}. 

Let us conclude by acknowledging that we have considered works in which the dynamics is well-stirred in the nodes, while diffusion or other interactions take place through the links. Indeed, there are some works in which the network describes the interactions between the species, e.g., see Zheng et al. \cite{zheng2016identifying} and Diego et al. \cite{diego2018key}. A discussion of this framework goes beyond the scope of this work.

\section{{  Master Stability Fuction theory}: coupled nonlinear systems approach}\label{sec:2nd_approach}

In the previous sections, we have considered the case where Turing patterns emerge from the perturbation of a stationary equilibrium solution. A step forward to relax this assumption has been provided by Challenger et al.~\cite{chall} by studying the Turing instability originated from a limit cycle solution. It is worth noticing that the analysis developed in ~\cite{chall} employs techniques made popular in the study of the synchronization of coupled nonlinear oscillators~\cite{arenas2008synchronization}{  , where diffusion-driven instabilities can occur as well and are called Benjamin-Feir instabilities \cite{nakao2009diffusion}. Synchronization} refers to the oscillatory collective behavior of systems {composed by} many constituents, documented for the first time by the Huygens in the 16th century \cite{pikovsky2001synchronization}, and successively formalized by Winfree \cite{winfree1967biological} and Kuramoto \cite{kuramoto1975}. Let us observe that synchronization can be achieved also by considering chaotic systems. Sufficient conditions for synchronization have been firstly proposed by Fujisaka and Yamada in a seminal work \cite{fujisaka1983stability} where they introduced the {\em stability parameter} and related it with some property of the support. The authors did not explicitly mention the use of networks, however in the current jargon we can say that they considered all-to-all coupling, a $2$-ring and a $1$-dimensional path. These ideas have been exploited and reformulated $15$ years later, by Pecora and Caroll, who introduced the formalism of the {\em Master Stability Function} \cite{pecora1998master}. As we will show in the next sections, tools and ideas used {  for} synchronization can be applied to the case of fixed point dynamics, leading to the same equations as those of reported in Sec. \ref{sec:1st_approach}. The equivalence between the two frameworks was known within the community of nonlinear physics, but it was rarely made explicit \cite{muolo_phd_thesis}. 

{  In the next section, we will show how a Turing-like instability can indeed emerge in systems of coupled oscillators, while the following section is dedicated to formally deriving the Master Stability Function and showing the equivalence between the formalism of coupled nonlinear systems, on the one side, and that of reaction-diffusion equations on networks, on the other.}

\subsection{Turing-like instability for oscillatory systems}

Let us start from a system of two interacting species, as in Eq. \eqref{eq:react}, but with the difference that now the homogeneous solution is a stable limit cycle with period $T$, i.e.,  $(u^*(t),v^*(t))$, with $u^*(t+T)=u^*(t)$ and $v^*(t+T)=v^*(t)$. Let us consider the above dynamics on a network of $n$ nodes, leading to a reaction-diffusion system with the same form of Eq. \eqref{eq:react_diff_eq}. As before, the local (oscillating) dynamics happens on the nodes, while Fickean diffusion takes place through the links. Such a setting has been first considered by Challenger et al. on networks in \cite{chall}, where it was shown that the homogeneous periodic solution can undergo a diffusion-driven instability  yielding pattern formation, hence analogous to the Turing one. Such results have been further corroborated by Van Gorder in \cite{vangorder} on a continuous support.

Let us proceed by perturbing the homogeneous {periodic} state with an inhomogeneous perturbation $\vec{\zeta}$ and then linearize {the system about such a reference solution}. One can proceed as in Sec. \ref{sec:1st_approach}, but with the key difference that now the linearized functions are $T$-periodic, i.e., $f_u(t+T)=f_u(t)$ and analogously for $f_v,~g_u,~g_v$, which means that the Jacobian is also periodic. The perturbation {  $\vec{\zeta}=(\delta u_1,\dots,\delta u_n,\delta v_1,\dots, \delta v_n)^\top$} evolves according to the following equation
\begin{equation}
\label{eq:2n2n_new}
\dot{\vec{\zeta}}=\left(\mathbf{J}_f(t) \otimes \mathbb{I}_n  +\mathbb{D}\otimes \mathbf{L} \right) \vec{\zeta}. 
\end{equation}
Let us observe that the above system is non-autonomous, the reason for which we have made explicit the time-dependency of the Jacobian, an information which will be omitted from now on when clear from the context. Exactly as before, we can project on the basis of the Laplacian, obtaining $n$ uncoupled systems for the evolution of the perturbation {  in the new basis}, namely, 
\begin{equation}
\dot{\vec{\xi}}_\alpha=\left[\mathbf{J}_f(t)+\Lambda^{(\alpha)}\mathbb{D}\right] \vec{\xi}_\alpha=\mathbf{J}_\alpha(t)\vec{\xi}_\alpha .
\label{eq:new_pert1_new2}
\end{equation}

Given the time dependence of the Jacobian, an analytical expression for the dispersion relation cannot be obtained, in contrast with the classic Turing setting in which the homogeneous solution is a fixed point. The system being periodic, {one can resort to the use of Floquet theory as} in Challenger et al. \cite{chall}, {where}  authors numerically computed the maximum Floquet exponent to obtain the modes for which the system would relax back to its stable limit cycle or, instead, yield Turing-like patterns. {Observe that Van Gorder used a complementary approach which relies on the oscillatory property of non-autonomous $2$-dimensional linear systems~\cite{vangorder}.} As we are about to show, such a method can be further generalized to the case in which the homogeneous solution is a chaotic attractor; one then needs to compute the maximum Lyapunov exponent as a function of the coupling, i.e., the celebrated Master Stability Function. Let us conclude by mentioning two results regarding non-normality. First, also in the context of a homogeneous periodic solution, it was found that a linear stability analysis may fail in predicting the final state when the underlying network is non-normal \cite{entropy}; moreover, a structural property of non-normal networks, i.e., the presence of nodes with only out-going or in-coming links \cite{asllani_leaders}, induces chimera patterns \cite{muolo2024persistence}.

\subsection{Derivation of the Master Stability Function}

{The analysis performed in the previous section yields, in the case of a stationary reference solution, the dispersion relation, that ultimately rests on two reference pillars: linearization and projection onto a suitable basis. A similar idea leads to the {   formalism developed by Fujisaka and Yamada \cite{fujisaka1983stability}, later reformulated and renamed {\em Master Stability Function} (MSF) by Pecora and Carroll \cite{pecora1998master},} to study synchronization of coupled nonlinear systems.}
The aim of this section is to introduce the reader to the formalism of the Master Stability Function for the setting of coupled nonlinear systems. Consider thus an ensemble of $n$ identical $m$-dimensional units whose state is described by the vector $\vec{x}_i\in\mathbb{R}^m$, coupled through a network. The equations describing the evolution of the system are given by
\begin{equation}\label{eq:synch}
    \dot{\vec{x}}_i=\vec{f}(\vec{x}_i)+\sigma \sum_{j=1}^n A_{ij} \vec{l}(\vec{x}_j,\vec{x}_i)~,~~\forall i=1,2,\dots,n,
\end{equation}
where $n$ is the number of nodes of the network, $\vec{f}$ a nonlinear vectorial function describing the dynamics of the decoupled system, $\sigma$ the coupling strength and $\vec{l}$ a vectorial function describing the coupling between the units through the network topology. Because we are dealing with identical systems, i.e., $\vec{f}$ does not depend on the node index, and because we assume $\vec{l}$ to be non-invasive, i.e., $\vec{l}(\vec{x},\vec{x})=0$ for all $\vec{x}$, we can conclude that a solution of the isolated system, $\dot{\vec{x}}^*(t)=\vec{f}(\vec{x}^*(t))$, is also a homogeneous solution of the whole system, {   even if} we take into account the coupling. Let us observe that we hereby assume, as often done in the literature, the non-invasive coupling to be realized by using the diffusive-like term, i.e., $\vec{l}(\vec{x}_j,\vec{x}_i)=\vec{h}(\vec{x}_j)-\vec{h}(\vec{x}_i)$ for some function $\vec{h}$. Global synchronization is thus realized whenever the above homogeneous solution is stable: in such a way, tiny perturbations fade away and the whole system converges back to the unperturbed solution, characterized by a coordinated set of trajectories evolving in unison, e.g., $\vec{x}_i(t)-\vec{x}^*(t)\rightarrow 0$ as $t\rightarrow \infty$ for all $i$. In the following, we will be interested in answering this question by taking into account dynamical factors, i.e., given by $\vec{f}$, and the coupling encoded by the network and the function $\vec{h}$. The stability of the synchronous solution can be studied through the formalism of Fujisaka and Yamada \cite{fujisaka1983stability} and the Master Stability Function \cite{pecora1998master}.

In what follows, we will elabrate on the key points of the theory, in order to highlight the analogies between this framework and the Turing setting. Let us perturb system \eqref{eq:synch} about the synchronous solution and linearize {it}. {  By denoting the variables of the $i$-th $m$-dimensional system as $\vec{x}_i=(x_i^{(1)},\dots,x_i^{(m)})$, i.e., $x_i^{(j)}$ is the $j$-th variable (species) on the $i$-th node, we define the perturbation about the reference solution as \begin{displaymath}  \vec{\zeta}=(\delta x_1^{(1)},\dots,\delta x_n^{(1)},\delta x_1^{(2)},\dots, \delta x_n^{(2)},\dots,\delta x_1^{(m)},\dots,\delta x_n^{(m)})^\top. \end{displaymath} The latter evolves according to}
\begin{equation}
\label{eq:lnln}
\dot{\vec{\zeta}}=\left( \mathbf{J}_f \otimes\mathbb{I}_n +\mathbf{J}_h \otimes \mathbf{L}\right) \vec{\zeta} , 
\end{equation} where now the perturbation vector $\vec{\zeta}$ has dimension $nm$ and the $m\times m$ matrices $\mathbf{J}_f$ and $\mathbf{J}_h$ are the Jacobian of functions $\vec{f}$ and $\vec{h}$, respectively, {evaluated on the solution $\vec{x}^*(t)$ and thus} they will be, in general, time-dependent. The network Laplacian appears due to the hypothesis of diffusive-like inter-nodes couplings. To facilitate analytical progress in the study of Eq.~\eqref{eq:lnln}, we must operate a change of basis to decompose the perturbation along different modes; from the above $nm\times nm$ system, we then obtain $n$ systems of dimension $m\times m$. A suitable basis is the one that diagonalizes the Laplacian {  matrix} $\mathbf{L}$, which allows to rewrite Eq. \eqref{eq:lnln} as
\begin{equation}
\dot{\vec{\xi}}_\alpha=\left[\mathbf{J}_f+\Lambda^{(\alpha)}\mathbf{J}_h\right] \vec{\xi}_\alpha=\mathbf{J}_\alpha\vec{\xi}_\alpha,
\quad \forall \alpha=1,\dots,n \, ,
\label{eq:new_pert}
\end{equation} 
where, again, $\Lambda^{(\alpha)}$ is the $\alpha$-th eigenvalue of $\mathbf{L}$. The first mode, associated to the $\Lambda^{(1)}=0$ eigenvalue, {describes} the perturbation parallel to the synchronization manifold. All the other modes are {transversal} to the latter and the computation of the MSF 
allows to identify the subset of those which can drive the system away from the synchronous state.

{The reader can thus appreciate the similarity of the MSF formalism and the concept of a dispersion relation as presented in Sec. \ref{sec:1st_approach}, the main difference being that Turing instability looks for the condition to turn unstable an otherwise stable stationary equilibrium, whereas the MSF determines the conditions for the stability of a generic reference solution, that can be stationary, time-dependent, but also chaotic.} 
{Those differences manifest themselves in the depth of the analysis one can perform.} In the case of fixed point the MSF can be computed analytically, while for other classes of solutions, one needs in general to resort to numerical simulations, since the linearized system is non-autonomous.  Remarkable exceptions are however possible \cite{nakao2014complex}.

\section{From pairwise to higher-order interactions}\label{sec:higher}

{Despite the major advances that network sciences allowed to the study of dynamical systems, among which Turing instability and synchronization,} scholars have realized over the past years that {  such a simple} and effective formalism hides a big limitation: in fact, it captures only pairwise interactions among the elementary units  {forming the system}. Many natural and engineered systems exhibit also group interactions, which may be of a different nature than pairwise ones. For example, co-authorship networks are intrinsically higher-order, as they are made of group interactions, i.e., co-authoring a scientific publication,
involving, in general, more than two authors \cite{patania2017shape}. {   Further evidence comes from neuroscience \cite{rosen1989,wang2013,petri2014homological,sizemore2018cliques}, ecology \cite{grilli_allesina} and social behaviors of humans \cite{centola2018experimental} and animals \cite{iacopini2024not}.} Some of the examples mentioned above are not recent, and such concepts were already widely accepted in those disciplines, while the community of dynamics on networks has only recently turned its interest towards them, with few exceptions {  \cite{tanaka2011multistable,krawiecki2014chaotic,ashwin2016hopf,bick16}}.
{To go beyond pairwise interactions, one should consider \textit{higher-order} (or \textit{many-body}) interactions, where more than two basic units interact at the same time. This can be modeled by using} hypergraphs \cite{berge1973graphs} and simplicial complexes \cite{aleksandrov1998combinatorial}. The former were developed mostly in the framework of computer science, while the latter by physicists and mathematicians. 
For what concerns the dynamics, many-body interactions naturally emerge once applying a center manifold analysis \cite{kuramoto2019concept}, as shown by Kuramoto already in 1984 \cite{Kuramoto_book} for continuous media{  ,} then by Kori et al. for globally coupled oscillators \cite{kori2014clustering} and by Contemori et al. for systems on directed networks~\cite{PhysRevE.93.032317}. A similar result has been provided by León and Pazó \cite{leon19} while performing phase reduction, a powerful technique used to obtain a phase description of systems of weakly coupled oscillators \cite{nakao16}. However, in both these settings, such interactions are weak and {  do not have much effect of the dynamics. On the other hand, the situation is different when those terms are strong: in fact}, scholars investigated different settings and phenomena, such as random walks \cite{schaub2020random,carletti2020random,carletti2021random}, consensus
\cite{neuhauser2020multibody,neuhauser2021}, synchronization of phase models \cite{skardal2019abrupt,lucas2020multiorder,leon2024,bick_explosive,skardal2019higher}, synchronization of chaotic oscillators \cite{gambuzza2021stability}, phase transitions \cite{de2021phase}, epidemics \cite{stonge2021universal}
and social contagion \cite{iacopini2019simplicial,de2019social}, to name a few, and found that many-body interactions dramatically affect the global behavior of the system. The same is true for Turing pattern formation, which has been studied in systems with higher-order interactions in a few works \cite{carletti2020dynamical,gao2023turing,muologallo, wang2023spatio}, as we will discuss in the following sections. Let us point out that, while \cite{carletti2020dynamical,gao2023turing,muologallo} refer to the classic "fixed point" Turing setting, the work of Wang and Liu \cite{ wang2023spatio} treats the case of non-autonomous systems, discussed in the previous section, with higher-order interactions. {   Let us also mention triadic interactions, where a node can regulate the interactions between  two others \cite{sun2023dynamic}, with interesting effects on the dynamics \cite{millan2023triadic}.}

{  In the next section, we will first introduce some basic concepts of higher-order structure, which will be then applied to the  study of the generalized Turing instability.  The same formalism can also be used to analyze dynamics beyond Turing patterns, such as synchronization, which will not be discussed here. The} interested reader can consult the following reviews \cite{battiston2020networks,natphys,boccaletti2023structure,gao2023dynamics,bick2023higher} and books \cite{bianconi2021higher,battiston2022higher} for further insights.

\subsection{Introduction to higher-order structures}

As seen in Sec. \ref{sec:1st_approach}, pairwise interactions are encoded through the adjacency matrix $A$,  {$A_{ij}=1$ if nodes $i$ and $j$ are connected}  \cite{newmanbook2}. In the higher-order framework, one needs \textit{adjacency tensors} to fully capture the interactions. A $(d+1)$-body interaction will be represented through the $d$-th order adjacency tensor $\textbf{A}^{(d)}$: 
$A^{(d)}_{ij_1...j_d}=1$ if there is a $(d+1)$-body interaction between nodes $i,j_1,...,j_d$. In such case, the higher-order structure  {is encoded by} a \textit{hyperedge}  {containing} nodes $i,j_1,...,j_d$.  {Because the hyperedge is an unordered collection of nodes,} the adjacency tensors are symmetric, {namely, $A^{(d)}_{\pi(ij_1...j_d)}=1$ for any permutation $\pi$ of the $(d+1)$ indexes  {forming the hyperedge}}. In light of this new definition, we can observe that the adjacency matrix is a $1$-st order adjacency tensor, since it encodes the $2$-body interactions, and can be referred to as $\textbf{A}^{(1)}$. We can also generalize the concept of degree, by defining the $d$-\textit{degree} $k_{i}^{(d)}=\frac{1}{d!}\sum\limits_{j_1,..,j_d=1}^N A_{ij_1\dots j_d}^{(d)}$, indicating the number of hyperedges of order $d$ of which node $i$ is part of. {   In a hypergraph, the presence of a $d$-body interaction does not imply that there are also lower order interactions, while a simplicial complex is closed by inclusion, i.e., if a $d$-body interaction is present, so are all the corresponding $(d-1)$-body interactions, and so on for lower orders.} 
Let us point out that, in the literature, scholars tend to refer to higher-order interactions as \textit{hyperedges}, when the support is a hypergraph, and \textit{simplices} (or \textit{simplexes}) when it is a simplicial complex. Moreover, in some references, a $d$-body interaction is called $d$-hyperedge, while in others it is called a $(d-1)$-simplex. The latter definition is used because in the theory of simplicial complexes nodes are called $0$-simplices, links ($2$-body) are $1$-simplices, and so on. In this Review we will use the latter convention, hence a $d$-body interaction will be a $(d-1)$-hyperedge or, if the closure relation is satisfied, a $(d-1)$-simplex. 

In this Section, we will consider the (mean-field) dynamics as taking place within the nodes, while the higher-order structures model the interactions among the units. Hence, this is a straightforward extension of the pairwise setting. Let us observe that in this way we can deal with discrete support which cannot be obtained as discretized manifolds.

\subsection{Turing patterns on higher-order structures}

Turing theory has been first extended to {hypergraphs} in~\cite{carletti2020dynamical} {where authors have used a sort of hyperedge mean-field: namely, the action of the nodes inside a hyperedge toward a given node, also belonging to the same hyperedge, is modulated by a function of the hyperedge size. This approach is equivalent to introducing a suitable weighted network and studying the Turing patterns on the latter. Successively a similar idea has been developed in~\cite{gao2023turing} to the framework of simplicial complexes. Eventually, a general theory of Turing patterns on higher-order structures has been proposed in~\cite{muologallo} based on the formalism developed by Gambuzza et al. \cite{gambuzza2021stability} to deal with synchronization {  dynamics}}. In the following, we will thus present this approach.

{   Let us consider a system of two interacting species, $u$ and $v$, defined on the $n$ nodes of a hypergraph of order $P$, namely the largest hyperedge size is given by $P+1$. We require the coupling functions (i.e., the terms that generalize the linear diffusion ones) to be nonlinear, so as to deal with actual many-body dynamics. In fact, if the coupling functions were linear, a $3$-body interaction could be reduced to three $2$-body ones, by introducing a suitable weighted adjacency matrix~\cite{neuhauser2020multibody}.} The interacting species status are described  {in each node} by the vector $\vec x_i=(u_i,v_i)^\top$, the reaction dynamics is given by $\vec{f}(\vec{x})=(f(u,v),g(u,v))^\top$ and {  the coupling by $\vec{h}^{(d)}(\vec{x}_{i},\vec{x}_{j_1},\dots,\vec{x}_{j_d})=({h}^{(d)}_1(u_{i},u_{j_1},\dots,u_{j_d},v_{i},v_{j_1},\dots,v_{j_d}),{h}^{(d)}_2(u_{i},u_{j_1},\dots,u_{j_d},v_{1},v_{j_1},\dots,v_{j_d}))^\top$. Since our aim is to discuss Turing patterns in higher-order reaction-diffusion systems, let us  assume the latter to be diffusive-like, namely, for $s=\{1,2\}$ 
\begin{displaymath}
{h}^{(d)}_s(u_{i},u_{j_1},\dots,u_{j_d},v_{i},v_{j_1},\dots,v_{j_d})=h_s^{(d)}(u_{j_1},\dots,u_{j_d},v_{j_1},\dots,v_{j_d})-h_s^{(d)}(u_i,\dots,u_i,v_i,\dots,v_i).
\end{displaymath}} For all $d=1,\dots, P$, the higher-order reaction-diffusion dynamics of the two species is described by the following system
\begin{equation}
\label{eq:sec5:react-diff}
 \begin{cases}
 \displaystyle 
 \dot{u}_i=f(u_i,v_i)+\sum_{d=1}^P\sigma_d\sum_{j_1=1}^n \dots \sum_{j_d=1}^n A_{i,j_1,\dots,j_d}^{(d)}\bigg[h_1^{(d)}(u_{j_1},\dots,u_{j_d},v_{j_1},\dots,v_{j_d}) -h_1^{(d)}(u_i,\dots,u_i,v_i,\dots,v_i)\bigg], 
 \\ \\ \displaystyle \dot{v}_i=g(u_i,v_i) + \sum_{d=1}^P\sigma_d\sum_{j_1=1}^n \dots \sum_{j_d=1}^n A_{i,j_1,\dots,j_d}^{(d)}\bigg[h_2^{(d)}(u_{j_1},\dots,u_{j_d},v_{j_1},\dots,v_{j_d})   -h_2^{(d)}(u_i,\dots,u_i,v_i,\dots,v_i)\bigg].
 \end{cases}
\end{equation}
Let us assume that the nonlinear diffusion does not contain any cross-diffusion term, namely for all $d\in\{1, \dots, P\}$, the function $h_1^{(d)}$ (resp. $h_2^{(d)}$) depends only on $\{u_{j_1},\dots,u_{j_d}\}$ (resp. $\{v_{j_1},\dots,v_{j_d}\}$) and let us {  make} explicit the diffusion coefficients of each species at every order, namely, $h_1^{(1)}\mapsto D_u^{(1)} h_1^{(1)}$ and analogously for $D_v^{(1)},~D_u^{(2)}, D_v^{(2)}$, and so on. The theory is, of course, general for any order, but let us further simplify the analysis by considering only pairwise and $3$-body interactions (i.e., $P=2$). The equations for the higher-order reaction-diffusion system become 
\begin{equation}
\label{eq:sec5:react-diff2}
 \begin{cases} 
 \displaystyle \dot{u}_i=f(u_i,v_i)+\sigma_1 D_u^{(1)}\sum_{j_1=1}^n A_{ij_1}^{(1)}(h^{(1)}_1(u_{j_1})-h^{(1)}_1(u_i)) +\sigma_2 D_u^{(2)}\sum\limits_{j_1=1}^n\sum\limits_{j_2=1}^n A_{ij_1 j_2}^{(2)}(h^{(2)}_1(u_{j_1},u_{j_2})-h^{(2)}_1(u_i,u_i)),\\ \\ \displaystyle \dot{v}_i=g(u_i,v_i) + \sigma_1 D_v^{(1)}\sum_{j_1=1}^n A_{ij_1}^{(1)}(h^{(1)}_2(v_{j_1})-h^{(1)}_2(v_i)) +\sigma_2 D_v^{(2)}\sum\limits_{j_1=1}^n\sum\limits_{j_2=1}^n A_{ij_1 j_2}^{(2)}(h^{(2)}_2(v_{j_1},v_{j_2})-h^{(2)}_2(v_i,v_i)).
 \end{cases}
\end{equation}

Let us observe that the coefficients $\sigma$ are pleonastic and, in principle, could be absorbed into the diffusion coefficients. However, it is easier to capture the effects of higher-order interactions by keeping the diffusion coefficients of the same order of magnitude and tuning the strength of a given interaction through the parameter $\sigma$.

We can now repeat the same linear stability analysis of Sec. \ref{sec:1st_approach}, by assuming the existence of a stable homogeneous solution $(u^*,v^*)$ and expand an inhomogeneous perturbation up to the first order. {{   The presence of higher-order interactions poses some challenges, which have been elegantly overcome by Gambuzza et al. in \cite{gambuzza2021stability}. We will skip the details of the analysis, that the interested reader can find in the aforementioned reference, and we report here the main conclusions.} Let $\vec{\zeta}=(\delta u_1,\delta v_1,\dots,\delta u_n,\delta v_n)^\top$, where $\delta u_i=u_i-u^*$ and $\delta v_i=v_i-v^*$, be the perturbation vector, {  $\mathbf{J}_f=\left(
\begin{smallmatrix}
 f_u & f_v\\
 g_u & g_v
\end{smallmatrix}
\right)$ the Jacobian matrix of the isolated system}, $\mathbf{J}_{h}=\mathbf{J}_{h}^{(1)}$ and  $\mathbf{J}_{h}^{(2)}=\left(
\begin{smallmatrix}
 \partial_{u_1} h^{(2)}_1+\partial_{u_2} h^{(2)}_1 & 0\\
 0 & \partial_{v_1} h^{(2)}_2+\partial_{v_2} h^{(2)}_2
\end{smallmatrix}\right)$, where the derivatives are evaluated on the homogeneous solution. The evolution of the {  perturbation {  $\vec{\zeta}=(\delta u_1,\dots,\delta u_n,\delta v_1,\dots, \delta v_n)^\top$}} is then given by
\begin{equation}
\label{eq:sec5:general_high_order_lin_eq}
\dot{\vec{\zeta}}=\left(\mathbf{J}_f \otimes \mathbb{I}_n +\sigma_1 \mathbf{J}_{h}^{(1)}\otimes\mathbf{L}^{(1)} +\sigma_2 \mathbf{J}_{h}^{(2)} \otimes \mathbf{L}^{(2)} \right) \vec{\zeta}, 
\end{equation}
{   with $\mathbf{L}^{(1)}$ the usual network Laplacian and $\mathbf{L}^{(2)}$ its generalization \cite{gambuzza2021stability} accounting for the three-body interactions, defined as 
\begin{equation}
\label{eq:2-laplacian}
\mathbf{L}^{(2)}_{ij} = \begin{cases}
-\sum_{j,k=1}^{N} A^{(2)}_{ijk} & \mathrm{for} \ i=j, \\ \\
\phantom{+} \sum_{k=1}^{N} A^{(2)}_{ijk} & \mathrm{for} \ i\neq j. \\
\end{cases}
\end{equation}}

When considering general coupling functions and higher-order topologies, the above equation cannot be diagonalized, because the terms $\mathbf{J}_{h}^{(1)}\otimes\mathbf{L}^{(1)}$ and $\mathbf{J}_{h}^{(2)} \otimes \mathbf{L}^{(2)}$ do not commute. This means that one {cannot obtain} an analytical expression for the dispersion relation. The emergence of Turing patterns can hence only be studied numerically {being, in general, very difficult, if not impossible, to obtain an explicit expression for the eigenvalue with the largest real part}. As discussed in \cite{muologallo}, one can deal with cases in which the first-order diffusion coefficients (i.e., pairwise) do not allow for the Turing instability, but the presence of higher-order diffusion with a suitable choice of the parameter $\sigma_i$ can yield the formation of patterns. \textit{Vice versa}, {  in other cases} Turing patterns would arise in the pairwise setting, but many-body interactions can suppress the instability and stabilize the system. 

Let us now examine two special cases in which the dispersion relation can be obtained analytically. {   We start by observing that the Laplacians $\mathbf{L}^{(1)}$ and $\mathbf{L}^{(2)}$ are symmetric and zero-row-sum matrices, diagonalizable and with a real negative semi-definite spectrum, to which corresponds an orthonormal basis. However, {they cannot be simultaneously diagonalized and, thus, we cannot disentangle the dynamics originating from different modes. To overcome this issue}, authors of \cite{gambuzza2021stability} {introduced} the hypothesis of \textit{natural} coupling, namely, $\forall \vec{x}\in\mathbb{R}^m$
\begin{equation*}
\vec{h}^{(d)}(\vec{x},\dots,\vec{x})=...=\vec{h}^{(2)}(\vec{x},\vec{x})=\vec{h}^{(1)}(\vec{x}).
\end{equation*} Physically, this assumption means that all coupling functions have the same effect when the whole system lies in a homogeneous state. In our case, such hypotesis implies that 
\begin{equation*}
{h}_1^{(2)}(u,u)={h}_1^{(1)}(u)\text{ and }{h}_2^{(2)}(v,v)={h}_2^{(1)}(v) ~~~\Rightarrow ~~~ \mathbf{J}_{h}^{(1)}=\mathbf{J}_{h}^{(2)} \equiv\mathbf{J}_{h} .
\end{equation*}}

If we set $h^{(1)}_1(u) = D^{(1)}_u u^3$ and $h^{(1)}_2(v) = D^{(1)}_v v^3$ for the pairwise coupling and $h^{(2)}_1(u_1,u_2)=D^{(2)}_u u_1^2u_2$ and $h^{(2)}_2(v_1,v_2)=D^{(2)}_v v_1^2v_2$ for the three-body interactions, {  the above hypothesis implies that} the diffusion coefficients are the same for every order, namely, $D_u^{(1)}=D_u^{(2)} \text{ and } D_v^{(1)}=D_v^{(2)}$. Now, Eq. \eqref{eq:sec5:general_high_order_lin_eq} becomes \begin{equation}
\label{eq:natural_lin}
\dot{\vec{\zeta}}=\left[ \mathbf{J}_f \otimes\mathbb{I}_n + \mathbf{J}_{h}\otimes (\sigma_1\mathbf{L}^{(1)} +\sigma_2  \mathbf{L}^{(2)} ) \right] \vec{\zeta} .
\end{equation}

We can proceed by diagonalizing $\sigma_1\mathbf{L}^{(1)} +\sigma_2  \mathbf{L}^{(2)}$ and project, as before, the $2n\times 2n$ linear system to obtain $n$ linear systems of size $2\times 2$ depending each one on a single eigenvalue, $\Lambda^{(\alpha)}$, of $\sigma_1\mathbf{L}^{(1)} +\sigma_2  \mathbf{L}^{(2)}$, obtaining {  
\begin{equation}\label{eq:mse_natcoupl}
 \dot{\vec{\xi}}_{\alpha}=\left[\mathbf{J}_f+\Lambda^{(\alpha)}\mathbf{J}_{h}\right] \vec{\xi}_{\alpha}.
\end{equation}}

\begin{figure}
\centering
\includegraphics[width=\textwidth]{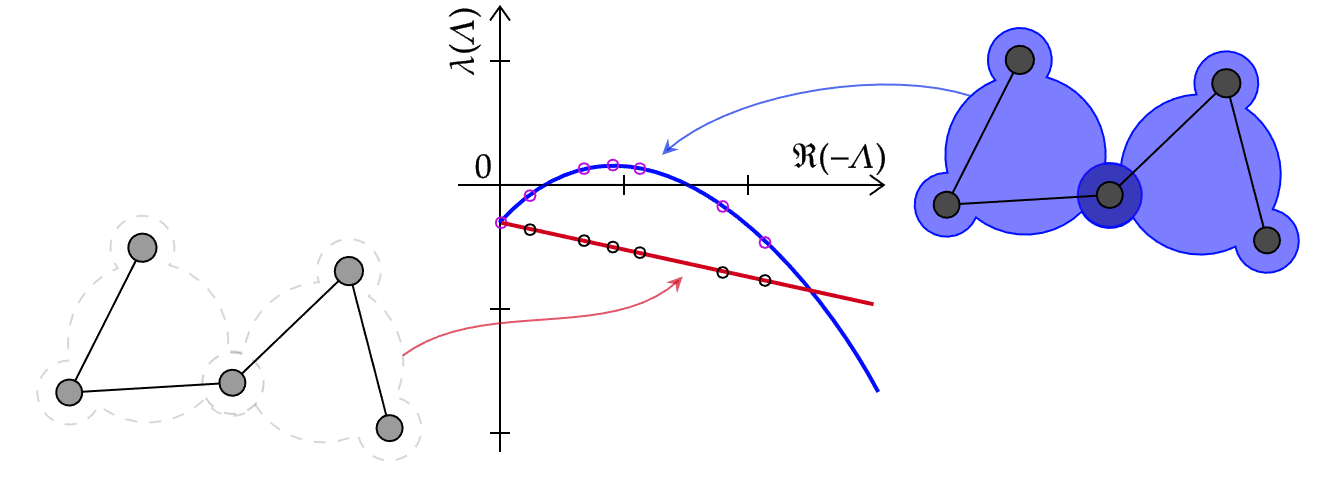}
\caption{{  The pairwise (network) setting would not allow for Turing instability, as the resulting dispersion relation (red curve) is negative, but the presence of higher-order interactions (in this case, $3$-body interactions)} with the appropriate diffusion coefficients allows the dispersion relation (blue curve) to take positive values and hence the higher-order system exhibits patterns. {  From Eq. \eqref{eq:mse_trig}, it can be appreciated that both the pairwise and higher-order dispersion relation have the same eigenvalues $\Lambda^{(\alpha)}$. Inspired by Fig. 3 of \cite{muologallo}, where the model is Brusselator \cite{PrigogineNicolis1967}.}}
\label{fig:3}
\end{figure}

The above case is equivalent to that discussed by Carletti et al. \cite{carletti2020dynamical}, Gao et al. \cite{gao2023turing} and 
 Wang and Liu \cite{wang2023spatio}. Here, {   the many-body interactions solely modify the $\Lambda^{(\alpha)}$, while the dispersion relation is not affected; stated differently, this is the same effect we would obtain by modifying the weights of a network. There} is however an effect due to large hyperedges where patterns tend to localize~\cite{carletti2020dynamical}. Alternatively, one can look for hypergraphs such that at least two generalized Laplacians are {simultaneously} diagonalizable. {A notable example is} the \textit{triangular lattice with periodic boundary conditions}, i.e., a $2$-torus paved with triangles, is such that $\mathbf{L}^{(2)}=2 \mathbf{L}^{(1)}$ (see \cite{muologallo} for details). Eq.~\eqref{eq:sec5:general_high_order_lin_eq} now takes the form
\begin{equation}
\label{eq:natural_lin_tri}
\dot{\vec{\xi}}=\left[\mathbf{J}_f \otimes \mathbb{I}_n +  (\sigma_1 \mathbf{J}_{h^{(1)}}+2\sigma_2 \mathbf{J}_{h^{(2)}}) \otimes \mathbf{L}^{(1)} \right] \vec{\xi} .
\end{equation}
{  Finally}, we can project on the eigenvectors of $\mathbf{L}^{(1)}$, obtaining
\begin{equation}\label{eq:mse_trig}
  \dot{\vec{\xi}}_{\alpha}=\left[\mathbf{J}_f+\Lambda^{(\alpha)}(\sigma_1 \mathbf{J}_{h^{(1)}}+2\sigma_2 \mathbf{J}_{h^{(2)}})\right]  \vec{\xi}_{\alpha} ,
\end{equation} {  which allows} us to analytically compute the dispersion relation. Let us remark that the above result can be straightforwardly generalized to all topologies yielding commutation relations between their Laplacians. Such topologies, called \textit{regular topologies} in \cite{muologallo}, are very specific and {maybe not suitable} for {real-world} applications; however, they provide a pedagogical case study to visualize the effects of higher-order interactions on the Turing mechanism. In Fig. \ref{fig:3}, we show the phenomenology of the mechanism {  for a triangular lattice with periodic boundary conditions}, with the higher-order interactions allowing the formation of patterns in a setting where the sole pairwise ones would not allow. Let us observe that also an the opposite situation can occur, namely, the pairwise setting allows the formation of patterns, which are suppressed by the higher-order interactions.

\section{Higher-order Turing patterns: topological signals and Dirac operator}\label{sec:top}

 {In the previous sections}, we have considered  {the setting where species are maant to react when sharing the same node and to diffuse across nodes by using the available links or hyperedges.} However, simplicial complexes, which are enriched by discrete topology structures, allow to further extend such framework  {by considering state variables (species) to be associated not only to nodes but also to links and higher-order structures, defining thus {\em topological signals}~\cite{bianconi2021higher,nakahara2003geometry,Lim2020}}. These include synaptic signals in neurons, edge signals in brain networks, and flows in biological transportation, power grids, and traffic networks \cite{linne2022neuron, faskowitz2022edges, santoro2023higher, katifori2010damage, rocks2021hidden, witthaut2022collective, barbarossa2020topological, sardellitti2022topological, schaub2018flow, schaub2021signal}.  {Once the coupling among signals of the same dimension is imposed through higher-order Laplacian operators,} they can undergo simplicial synchronization  \cite{millan2020explosive,millan2022geometry,torres2020simplicial,ghorbanchian2021higher,calmon2021topological,arnaudon2022connecting,deville2020consensus,reitz2020higher,ziegler2022balanced} and are analyzed by using topological signal processing and machine learning \cite{barbarossa2020topological,schaub2020random,schaub2021signal,bodnar2021weisfeiler,roddenberry2019hodgenet,hajij2020cell}. This field is recently burgeoning with novel lines of interesting research directions \cite{skardal2019abrupt,skardal2020higher,gambuzza2021stability,kovalenko2021contrarians,alvarez2021evolutionary,lee2021homological,carletti2020random,lucas2020multiorder,tang2022optimizing,zhang2021unified,chutani2021hysteresis}. 
 {Let us observe that one can also couple signals of different dimension by using the} Dirac operator, leading to phenomena like Dirac synchronization \cite{bianconi2021higher,josthorak2013spectra,bianconi2021topological,lloyd2016quantum,ameneyro2022quantum,knill2013dirac,calmon2021topological} and, as we will see, pattern formation \cite{turing_topological,muolo2024three}.

\subsection{{  Introduction to algebraic topology and topological signals}}\label{sec:geometric-view}

 {The aim of this section is to introduce the reader to  a selected gallery of basic concepts propaedeutic to the understanding of the extension of the Turing theory to topological signals; the interested reader is invited to consult Refs. \cite{Lim2020,bianconi2021higher,bianconi2021topological} for a broader and deeper introduction to the topic. Our goal here is to define some of the fundamental tools, namely the \textit{boundary} and \textit{coboundary operators} and the \textit{discrete Dirac operator}, which are the key to obtain topological Turing patterns}. The reader already familiar with these concepts can directly jump to the next section, where we  the theory of Turing pattern formation for topological signals is discussed.

An $n$-dimensional {\em simplex} is a set of $n+1$ nodes representing a higher-order interaction among ($n+1$) agents or a discrete space of dimension ($n+1$). For instance, a $0$-simplex is a node, a $1$-simplex a link, a $2$-simplex a triangle, and so on. The {\em faces} of a {  $d$-simplex} are formed by a subset of its {  $(d-1)$-simplices}. For example, the faces of a triangle ($2$-simplex) are its $3$ links {  1-simplices}, while each link has $2$ nodes as faces. In order to visualize simplicial complex we could start with a geometric picture. A simplex $\sigma_n$ in $\mathbb{R}^{n+1}$ is defined as:
\begin{equation}\label{geometrical_simplex}
	\sigma_n = \Big\{\left(t_0, \ldots, t_n\right) \in \mathbb{R}^{n+1} \mid \sum_{i=0}^n t_i=1 \text { and } t_i \geq 0 \text { for } i=0, \ldots, n\Big\}\, .
\end{equation}
In this representation, the simplex is a polytope with coordinates only between 0 and 1 that {  lie} in $\mathbb{R}^{n+1}$; for instance, with $n=2$, the simplex is a triangle with vertices located on the axis of {  $\mathbb{R}^{3}$}. We could imagine having several copies of this triangle and "gluing" them together via their links or vertices constructing a discrete manifold made by "cuts out" triangular pieces of {  $\mathbb{R}^{3}$}. By definition, when we glue two triangles together via vertices or edges, the latter belong to both involved triangles. 
A $d$-dimensional simplicial complex $\mathcal{S}$ includes simplices up to dimension $d$, glued together via their faces. This method of constructing a simplicial complex, although evocative and easy to visualize, is somewhat redundant. Many properties of the constructed structure can be described simply by noting which simplices are connected and through which faces.
We can therefore describe the structure by using a discrete notation and the proprieties obtained via combinatorial tools that act on the set of labels used to describe every simplex, and only preserving the closeness relations, i.e., all the faces of a simplex belong to the simplicial complex. This will be the starting point of the use of algebraic topology to describe a simplicial complex~\cite{bianconi2021higher, giamba_phd_thesis}. {  Let us point out that this definition should not be confused with the recently introduced Metaplexes \cite{estrada_2023}, where the simplices still preserve their continuous $n$-dimensional structure but are embedded in a graph discrete geometry. In order to properly define the structure by using only the labels, an (arbitrary) orientation needs to be set. This is necessary to determine the way higher-order objects such as links and triangles are glued together. For example, a link has a positive orientation from one node to another, and a triangle requires an orientation (defined, for instance, by the right-hand rule), which {can be} different from the orientation of its three links. In Figure \ref{fig:simplex} we represent a small $2$-simplicial complex composed by three triangles ($\sigma_2$), eight links ($\sigma_1$) and six nodes ($\sigma_0$). As the reader can appreciate, triangles are defined by using triplets of nodes, ordered according to the imposed orientation (see round arrows inside the triangles). Similarly, links are defined by using ordered pairs, the orientation being shown by the arrows on the triangles sides on the left panel. Observe that nodes can not be oriented. As previously stated, faces of each simplex can be described by using a combinatorial operator that acts on the list of vertices. As an example, let us consider Fig. \ref{fig:simplex}: the faces of the triangle $[2,4,3]$ are its three oriented links $-[4,2]$, $[4,3]$ and $[3,2]$; similarly, the faces of the link $[3,2]$ are the nodes $[2]$ and $[3]$. More formally, one can define the boundary operator acting on triangles, denoted by $\mathbf{B}_2$, that associates to each triangle its boundary; based on the above we get $\mathbf{B}_2[2,4,3]=[4,3]-[2,3]+[2,4]$, with the minus sign indicating the opposite orientation of the link with respect to the triangle side.} 

{  This construction can be generalized to a generic $n$-simplex represented via a list of n labels $[i_0, \dots, i_{n-1}]$ in the following way:
\begin{equation} \label{combinatorialBoundary}
   \bnd_n [i_0, \dots, i_{n-1}] = \sum_{k=0}^{n-1} (-1)^{k}[i_0, \dots,\hat{\imath}_k,\dots i_{n-1}]\, ,
\end{equation}
where we indicate by $\hat{\imath}_k$ the fact that the $k$-th label has been removed by the list. This way, information regarding the lower dimensional bounding simplices can be obtained by introducing the $n$-th boundary operator $\mathbf{B}_n$. Once a simplicial complex is given, we can therefore extract information regarding the boundary of each simplex at each level. For a given complex having $N_K$ simplices of dimension $k$ we can extract the set of $(k-1)$ simplices with the relative orientation, the latter being $N_{k-1}$ in number. All this information can be encoded in a $N_{k-1} \times N_k$ matrix that can be defined as: 
\begin{equation} \label{matrixBoundary}
    {\bf B}_k(i,j)=
    \begin{cases}
        1 & \text{ if } \sigma^{(i)}_{k-1}\sim \sigma_k^{(j)}\\
        -1 & \text{ if } \sigma^{(i)}_{k-1}\not\sim \sigma_k^{(j)}\\
        0 & \text{ otherwise},
    \end{cases}\,
\end{equation}
where $\sim$ indicates equal orientation and $\not\sim$ opposite.}

We will show that this operator maps functions defined on simplices onto lower-adjacent faces. For instance, {   when} acting on a link, it returns the difference of the signal at each extremity, i.e., {  the node faces}, operating similarly to a differential operator.

\begin{figure}
\centering
\includegraphics[width=\textwidth]{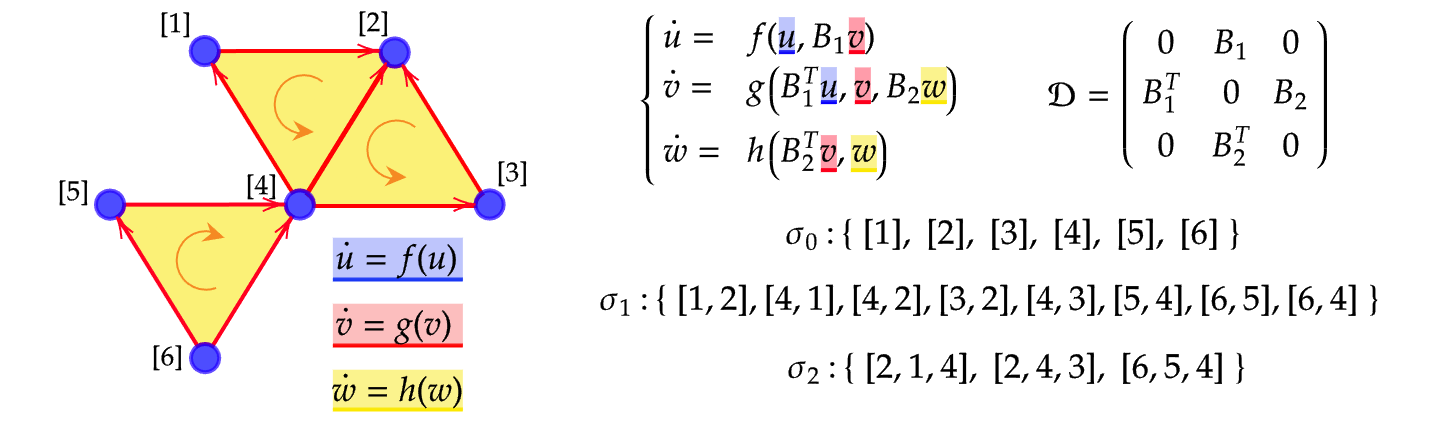}
\caption{{  An example of simplicial complex consisting of $0$-simplices $\sigma_0$ (nodes), $1$-simplices $\sigma_1$ (links) and $2$-simplices $\sigma_2$ (triangles). Let us remark that the structure is symmetric, while the orientation is needed to define the boundary and co-boundary operators, as explained in the text.} The three species (topological signals), $u$, $v$ and $w$, lie in different simplices and can interact via the Dirac operator $\mathcal{D}$.}
\label{fig:simplex}
\end{figure}

We can now intruduce the concept of \textit{topological signals}, which are functions defined on simplices such as {  nodes, links and triangles,} and are a natural extension of nodes dynamics. They can be understood as the discrete analogues of differential forms in differential topology, namely as vector fields, fluxes and so on, defined on discrete spaces. A $k$-dimensional topological signal $x$ can be conceptualized as a function $x$ mapping each $k$-dimensional simplex to a real number (it can be mapped to a real valued $d$-dimensional vector, but we will assume $d=1$ for simplicity). Represented as a vector $\vec{x}$ in a vector space of  dimension $N_k$, each component is $x_i = x(\sigma^{(i)}_k)$ for $\sigma^{(i)}_k \in S_k$, the latter being the set of all the $k$-dimensional simplices. 
The most relevant propriety of topological signal is the behavior with respect to the inversion of the simplex orientation: $x(\sigma^{(i)}_k) = -x(-\sigma^{(i)}_k)$. This needs to be satisfied to allow for invariance under arbitrary orientation of the simplices and has nontrivial consequences when functions over topological signals are considered.
Let $\vec{F}$ be a function transforming $\vec{x}$ to another topological signal {  $\vec{F}(\vec{x})=(f_1(x_1),\dots,f_{N_k}(x_{N_k})^\top$, namely, acting component-wise}. The latter must maintain orientation invariance, thus $f_i(x_i) = -f_i(-x_i)$. 
The dynamics is therefore described by {  $\dot{{x}}_i = f_i({x}_i)$,} where $f_i$ is a nonlinear, odd function to ensure invariance under simplex orientation reversal. The property of being an odd function is crucial for higher-order topological signals ($k>0$), which are vector-like and their interpretation depends on the orientation of the basis. {  Therefore, once we have given an arbitrary orientation to the higher dimensional simplices, we can define the components of a topological signal on the basis set by the orientation.} Every change of orientation will, however, rebound on the components with the minus sign, as stated above. Once the orientation is given for the higher dimensional set of simplices, a basis for the lower dimensional level can be obtained looking at the output of the boundary operator on every simplex. The output elements (without the sign coming from \eqref{combinatorialBoundary}) are the one to be used. Matrix \eqref{matrixBoundary} is the boundary operator on this basis of the two vectors spaces that can now be linked via its action.
To any given $k$-dimensional topological signal $\vec{x}$, one can associate a  $(k+1)$ signal $\vec{g}$~ by using the transpose of the incidence matrix $\vec{g} = \bnd_{k+1}^\top \vec{x}$ and a $(k-1)$ signal $\vec{z}$~ by using the incidence matrix $\vec{z} = \bnd_{k} \vec{x}$. {  With a great simplification, we can define the coboundary operator is the transpose of the boundary operator.} As we will see, different dimensional topological signals can be coupled via boundary and coboundary operators.

Now, starting from the definitions of boundary and coboundary operators, we can define the higher-order Laplacians and the Dirac operator.
The Laplacian operator of order $k$, also called $k$-Hodge-Laplacian \cite{josthorak2013spectra,Lim2020,bianconi2021higher}, describes higher-order diffusion from $k$-simplices to $k$-simplices and is an $N_k \times N_k$ matrix defined as 
\begin{equation}\label{hodgeLap}
	\mathbf{L}_{k}=\mathbf{B}_{k}^{\top} \mathbf{B}_{k}+\mathbf{B}_{k+1} \mathbf{B}_{k+1}^{\top}= \lap^{down}_k+\lap^{up}_k\, ,
\end{equation}
for $1\leq k<d$. For $k=0$ and $k=d$ the Hodge-Laplacians $\mathbf{L}_{0}$ and $\mathbf{L}_{d}$ are respectively given by $\mathbf{L}_{0} = \lap^{up}_0 = \mathbf{B}_{1} \mathbf{B}_{1}^{\top}$ and $\mathbf{L}_{d} = \lap^{down}_d = \mathbf{B}_{d}^\top \mathbf{B}_{d}$. Let us observe that this Laplacian is positive semi-definite, different from the Laplacian used in the previous sections, which was negative semi-definite; we have kept {   this definition as it is the one commonly used in such a context.} 

The term $ \lap^{up}_k $ represents the diffusion between $k$-simplices through shared $(k+1) $-dimensional simplices. In the case of a network, as previously noticed, this is the conventional {  network} Laplacian, where concentrations on nodes diffuse through incident links. The term $\lap^{down}_k$  represents diffusion between $k$-simplices through shared $(k-1)$-simplices, i.e., incident $(k-1)$-faces.
For instance, in a network (i.e., a $1$-simplicial complex), ${\bf L}_1={\bf L}_1^{down}$ determines diffusion from links to links through nodes. {  Referring to Fig. \ref{fig:simplex}, the topological signal defined on triangles $[2,1,4]$ and $[2,4,3]$ can be coupled via the $\lap^{down}$, as they share one link, but not via  $\lap^{up}$. However, none of the $2$-topological signals can be coupled with the one of triangle $[6,5,4]$.}
From this definition, it is clear that the Hodge-Laplacian of order $k$ only acts on topological signals of dimension $k$. Therefore, the $k$-Hodge-Laplacian cannot couple signals of different dimensions. {   This task is fulfilled by the discrete Dirac operator \cite{bianconi2021topological,lloyd2016quantum,ameneyro2022quantum}. In the above references, the reader can find a general definition, while we will focus on the case of networks, where the Dirac operator takes the following form}

\begin{equation}
	\mathcal{D}=\begin{pmatrix}0&{\bf B}_1\\
		{\bf B}_1^{\top}&0\end{pmatrix}.
\end{equation}
We can now understand the key property of the Dirac operator, i.e., that it can couple topological signals of different dimensions, differently from the Hodge-Laplacian. In particular, the Dirac operator can project a topological signal of any dimension $k$ onto simplices of dimensions $k+1$ and $k-1$.
One of the most remarkable properties of the Dirac operator is that it can be considered the ``square root" of the Laplacian. {   By denoting with $\oplus$ the direct sum, on networks we have $\mathcal{D}^2 = \mathcal{L} = \lap_0 \oplus \lap_1$.}

Let us now focus on the spectrum of the Hodge-Laplacians $\lap_0$ and $\lap_1$, and on the relation between their {  spectra} and the singular values of the boundary operator ${\bf B}_1$. 
Since $\lap_0={\bf B}_1{\bf B}_1^{\top}$ and $\lap_1={\bf B}_1^{\top}{\bf B}_1$, it follows that $\lap_0$ and $\lap_1$ are isospectral, i.e., they have the same non-zero eigenvalues and any eigenvalue $\Lambda_0^k$ of $\lap_0$ can be written as $\Lambda_0^k=b_k^2$ where $b_k$ indicates the non-zero singular values of the boundary matrix ${\bf B}_1$.
We note, however, that the degeneracy of the zero eigenvalue $\Lambda_0=0$ is different for $\lap_0$ and $\lap_1$. Indeed, for $\lap_0$ the degeneracy of the zero eigenvalue is the {  $0$-Betti number} $\beta_0$, i.e., the number of connected components of the network, while for $\lap_1$ it is given by the {  $1$-Betti number} $\beta_1$, i.e., the number of independent cycles of the network. Therefore, for a network that has the topology of a linear chain with periodic boundary conditions, we have $N_0=N_1$, and thus $\beta_0=\beta_1=1$; for a tree, when $N_1=N_0-1$, we have $\beta_0=1$ and $\beta_1=0$; in general, for a connected network, we have $\beta_0=1$, $\beta_1=N_1-N_0+1$.

\subsection{Patterns for Topological Signals}

{  This Section focuses} on reaction-diffusion systems in simplicial complexes, involving reaction, diffusion terms across simplices of the same dimension via the Laplacian matrix and interactions among simplices of different dimensions, by using the Dirac operator. For the sake of pedagogy, we hereby propose a Turing theory for topological signals in a simplicial complex where species lie on nodes {   and links; the system state is thus represented by a vector $\vec{\Phi}$ whose components $\vec{u}$ and $\vec{v}$, are the species concentrations on nodes and links, respectively. Let us observe that the framework can be generalized also to triangles and higher-order simplices.} Signal interaction occurs through projections on adjacent simplices, achieved by applying the Dirac operator $\mathcal{D}$ to $\vec{\Phi}$, forming projected signals $\vec{\Psi}=\mathcal{D}\vec{\Phi}$. Because the larger simplices we consider are {  links}, the Dirac operator, $\mathcal{D}$, is defined by using the incidence {  matrix ${\bf B}_1$, and its transpose}. The dynamical state encompasses both the signals $\vec{\Phi}$ and their projections $\vec{\Psi}$.
The reaction-diffusion process can be defined as:
\begin{equation}\label{top_sign_eq}
\dot{\vec{\Phi}} = \vec{F}(\vec{\Phi}, \mathcal{D}\vec{\Phi}) - \gamma\mathcal{L}\vec{\Phi} = \vec{F}(\vec{\Phi}, \vec{\Psi}) - \gamma\mathcal{D}\vec{\Psi} ,
\end{equation}
where \( \vec{F}(\vec{\Phi}, \mathcal{D}\vec{\Phi}) \) is a nonlinear function coupling signals of dimension \( k \) with those of dimensions \( k+1 \) or \( k-1 \), and \(\gamma\) is a diagonal matrix of diffusion coefficients. Again, let us observe that the minus sign in front of the diffusion term is due to the positive semi-definite definition of the Hodge Laplacian. For instance, {  in our case with $k=1$ we have 
\begin{equation}
\vec{F}(\vec{\Phi},\mathcal{D}\vec{\Phi})=\left(\begin{matrix} 
			\vec{f}_0(\vec{u}, \mathbf{B}_1 \vec{v}) \\
			\vec{f}_1(\vec{v},\mathbf{B}_1^{\top} \vec{u})  
			\end{matrix}\right),
\end{equation}}
\noindent where  $\vec{f}_{k}(\vec{x},\vec{y})$ are nonlinear functions, such that $\vec{f}_0(\vec{u}, \mathbf{B}_1\vec{v})=(f_0(u_1, (\mathbf{B}_1\vec{v})_1),\dots,f_0(u_{N_0}, (\mathbf{B}_1\vec{v})_{N_0})$, etc. Such a setting is depicted in Fig. \ref{fig:simplex}{  , where also triangles are considered}.
The matrix $\gamma$ in Eq. \eqref{top_sign_eq} is a diagonal matrix  {  
\begin{equation}
\gamma=\begin{pmatrix}\hat{D}_0&0\\
0&\hat{D}_1\end{pmatrix},
\end{equation}}

{  where $\hat{D}_k$ is the diagonal matrix of diffusion coefficients of each topological signal of order $k$, namely, $(\hat{D}_k)_{ij}=D^{(i)}_k \delta_{ij}$ where each species has its own diffusion coefficient $D^{(i)}_k$ with $i=1\dots N_k$.}

Therefore, Eq. \eqref{top_sign_eq} describes topological signals defined on the simplices of the simplicial complex that react with the projection of the topological signals defined in different dimensions while undergoing higher-order diffusion. Stated differently, we are coupling together scalar functions (node-defined) with flows (link-defined).
As already discussed, a stable homogeneous equilibrium is essential for the Turing mechanism, becoming unstable under certain values of the diffusion coefficients and topological conditions, leading to a heterogeneous state. In network-based dynamical systems with Laplacian coupling, the existence of this solution for the whole system necessitates the homogeneous vector, $(1,\dots,1)^\top$, to belong to the kernel of the network Laplacian. This condition is automatically met in a connected network, but the scenario changes in simplicial complexes. 
For topological signals, the existence of the homogeneous global solution requires the vector $(1,\dots,1)^\top$ to be in the kernel of the Dirac operator{  , i.e., $ \mathcal{D}(1,\dots,1)^\top=0$.} In the present setting, where we are dealing with nodes and links, this condition rewrites as {   $\mathbf{B}_1 (1,\dots,1)^\top=0$}. 
{  The above condition} corresponds to the requirement that each node has an equal number of in-going and out-going links  {with respect to the given orientation}. Stated differently{, we need the graph to be \textit{Eulerian}}. {  The extension of such condition to higher-order simplices is discussed in \cite{carletti2023global}\footnote{{  Moreover, let us note that weighted simplicial complexes \cite{baccini2022weighted} allow to overcome such constraint \cite{wang2024global}.}}}.   Hence,} by assuming to deal with an Eulerian graph, our goal is to derive the dispersion relation allowing us to determine the conditions for the Turing instability onset in the presence of a Dirac reaction term that couples the two topological signals of different dimensions. In a network, we have $\vec{\Phi}=(\vec{u},\vec{v})^{\top}$ and $\vec{F}(\vec{\Phi},\mathcal{D}\vec{\Phi})=\left(\vec{f}(\vec{u}, \mathbf{B}_1 \vec{v}),\vec{g}(\vec{v},\mathbf{B}_1^{\top} \vec{u})\right)^{\top}$ where $\vec{f}$ and $\vec{g}$ are two generic nonlinear functions, assumed to be applied component-wise on the vectors, i.e., $\vec{f}(\vec{u}, \mathbf{B}_1\vec{v})=(f(u_1, (\mathbf{B}_1\vec{v})_1),\dots,f(u_{N_0}, (\mathbf{B}_1\vec{v}))_{N_0})$. 
Here $\gamma$ reduces to the $(N_0+N_1)\times (N_0+N_1)$ block diagonal matrix with structure
\begin{equation}
\gamma = \begin{pmatrix}
D_0 \id_{N_0}& 0 \\
0 & D_1 \id_{N_1}
\end{pmatrix},
\end{equation}
where $D_0$ and $D_1$ indicate the diffusion coefficients of  {each} species defined on nodes and links respectively.
The  Dirac operator $\mathcal{D}$ and the Laplacian operator $\mathcal{L}$ are defined as the $(N_0+N_1)\times (N_0+N_1)$ matrices with block structure
\begin{equation}\mathcal{D}=\begin{pmatrix} 0 & \mathbf{B}_1 \\ \mathbf{B}_1^\top & 0 \end{pmatrix},\quad \mathcal{L}=\mathcal{D}^2=\begin{pmatrix}
    \mathbf{L}_0 & 0 \\
    0 & \mathbf{L}_1
\end{pmatrix}.
\end{equation}
It follows that Eq.(\ref{top_sign_eq}) can be explicitly rewritten as
\begin{equation}
\begin{cases}
\dot{\vec{u}}=\vec{f}(\vec{u}, \mathbf{B}_1\vec{v})-D_0\lap_{0} \vec{u}, \\
\dot{\vec{v}}=\vec{g}\left(\vec{v},\mathbf{B}_1^{\top} \vec{u}\right) - D_1\lap_{1} \vec{v}\, .
\end{cases}\label{eq:syst1}
\end{equation}

We can now repeat the linear stability analysis of Sec. \ref{sec:1st_approach} for system \eqref{eq:syst1}, i.e., assume the existence of a homogeneous equilibrium, impose its stability, then perturb it and linearize the system (see \cite{turing_topological} for the details). We obtain that Turing patterns can arise even with a single species living on the nodes and another species on the links and, moreover, the two species need to be both inhibitors, in contrast with the classical activator-inhibitor framework. Other interesting phenomenologies are the possibility of finding patterns with equal diffusion coefficients and that Turing instability of topological signals of a network will never be localized only on nodes or only on links, but will always involve both nodes and links signals.

Let us conclude by commenting on another interesting result of such a framework. Consider system \eqref{eq:syst1} with just Dirac couplings, i.e., the diffusion coefficients are set to zero.

It can be shown that the dispersion relation can turn positive, hence yielding  Dirac-induced patterns, i.e., patterns that are solely triggered by couplings encoded in the Dirac operator. Such a novel class of  patterns have been first found in \cite{turing_topological} and then thoroughly discussed in \cite{muolo2024three} for the cases of three interacting species, namely, two on the nodes and one on the links.

\section{Perspectives and future directions}

In this Review, we have discussed different approaches to the study of Turing patterns on networks and the recent results in the field, showing the ductility of the network approach and the possible interplay between the latter and the PDEs framework. Additionally, we thoroughly discussed possible extensions of Turing seminal ideas in a perspective where higher-order interactions are accounted for. We believe that the latter framework is particularly promising and we hope with this work to inspire further developments in this direction. For instance, directed hypergraphs have been considered to study the effects of directed higher-order interactions on synchronization \cite{gallo2022synchronization,della2023emergence,von2023hypernetworks}, and also directed simplicial complexes have been recently formalized \cite{gong2024higher}; both directed structures could be considered in the context of higher-order interactions. Also, the perspective of control could be a source of novel results, as pinning control for hypergraphs has been only recently developed \cite{de2022pinning} and so far Turing patterns have not yet been studied in such a sense. Still from a control point of view, higher-order interactions could have important applications in epidemics, as shown by Li et al. \cite{li2024increasing}. Moreover, we believe that the network and, in particular, the higher-order setting provides a natural embedding for the quantum framework, where the study of Turing patterns has just begun \cite{ardizzone2013formation,kato2022turing}. Besides all that, the possibilities offered by a network perspective to investigate the process of pattern formation are far from being thoroughly elucidated. For example, while Turing patterns have been studied on growing domains in the continuous setting \cite{krause2019influence,sanchez2019turing}, there are no results in this direction on networks or higher-order structures. Moreover, the framework of geometrical Turing patterns \cite{van2023emergence} has been only recently approached and, for instance, directed topologies have not yet been considered. As discussed in Sec. \ref{sec:extensions_network}, the effects of non-normal networks on the dynamics have not been completely understood and many open questions are still to be solved. Also, the interplay between a directed and a varying topology, the former enlarging and the latter shrinking the Turing region, would be another interesting direction to tackle. Pattern prediction and analysis pose several open problems that are starting to be understood only recently \cite{rogov2018pattern,rogov2019pattern} and a comprehensive account of the interplay between the network structure and the final pattern is still lacking. Lastly, the new framework of higher-order multiplexes \cite{chang2023combined,krishnagopal2023topology} provides another interesting setting in which Turing patterns have just started to be investigated \cite{ye2023pattern}.

Our aim here was to make of the Review  a pedagogical introduction to the field to scholars coming from the PDEs community. At the same time, we also aimed at  providing a general introduction to the framework of higher-order interactions for the researchers working in the field of Turing patterns. We hope that this effort will foster further collaboration and interplay between the variegated communities interested in the multifaceted Turing pattern applications.

\section*{Acknowledgements} R.M. and H.N. acknowledge JSPS, Japan KAKENHI JP22K11919, JP22H00516, and JST, Japan CREST JP-MJCR1913 for financial support. During part of this work, R.M. was supported by a FNRS-FRIA Fellowship, funded by the Walloon Region, Grant FC 33443. We are grateful to all the colleagues with whom we have worked and discussed over these past years on this topic. In particular, we would like to thank Malbor Asllani, Ginestra Bianconi, Francesca Di Patti, Mattia Frasca, Luca Gallo, Shigefumi Hata, Yuzuru Kato, Vito Latora, Alexander Mikhailov, Daniele Proverbio and Michael Schaub, whose feedback have been precious in the realization of this work.

\end{document}